\documentclass[11pt]{article}

\usepackage{amsthm}

\newtheorem{teo}{Theorem}
\newtheorem*{teon}{Theorem}

\newtheorem{prp}{Proposition}

\newtheorem{dfn}{Definition}

\newtheorem*{miracle}{Miracle}
\newtheorem*{miraclen}{One more Miracle}
\newtheorem{rem}{Remark}

\newtheorem*{exan}{Example}

\usepackage[margin=1 in]{geometry}

\usepackage{fixltx2e}
\MakeRobust{\overrightarrow}

\usepackage{mathtools}
\usepackage{bm}
\usepackage{tipa}
\usepackage{ amssymb} 

\usepackage{hyperref}
\usepackage{enumerate}
\usepackage{float}
\usepackage{xurl}
\usepackage{color}
\usepackage{graphicx}
\usepackage{amsmath}
 \usepackage{amssymb}
 \usepackage{empheq}
 \usepackage{bm}
 \usepackage{cancel}
 \usepackage{esvect}
 
 \newcommand{\jk}{J\hspace{-1.5mm}K}

 \newcommand{\R}{\mathbb{R}}

    \title{Conformally symplectic Chaplygin reduction in   rubber rolling of surfaces of revolution over the plane} \vspace{1mm}

   \author{ Jair Koiller\footnote{Instituto de F\'isica, Universidade Estadual do Rio de Janeiro, Brazil  (jairkoiller@gmail.com) }\\
      With an appendix by  Luis Garc\'ia-Naranjo\footnote{Departimento di Matematica ``Tulio Levi-Civita'', Universit\`a         di Padova (luis.garcianaranjo@math.unipd.it)}
   }

   \date{}

\begin{document}
 
 \maketitle 

 \vspace{-4mm}  
 
  \noindent 
  \begin{center}   
    To  the memory of Alexey V.  Borisov\footnote{This paper was submitted to special volume of Regular and
    Chaotic Dynamics dedicated to the legacy of A. V. Borisov with the title ``Comments on a paper about rubber rolling by  A. V. Borisov,   I. S. Mamaev and    I. A. Bizyaev".}.  \\
    Alexey did   Poetry through Mathematics  and Music. 
      Poetry  is not easy.
    \end{center}
 
\vspace{1mm}
{\large  \centerline{\bf{Abstract}} }

\vspace{3mm}

 {\small
\noindent    
   Rubber rolling (no-slip and   no-twist) of  a convex body on  the plane  under  the influence of 
  gravity
is  a  $SE(2)$ Chaplygin system, that reduces to $T^*S^2$,    the  sphere of Poisson vectors.  
I comment  upon
     an observation  by    
      A.V Borisov and I.S. Mamaev  (\cite{Borisov2008}, 2008) for   the case of  \textit{surfaces of revolution} [also in  
    A. V. Borisov,   I. S. Mamaev and    I. A. Bizyaev (\cite{Borisov2013}, 2013)].    They show that this  case  is  quite special:  \textit{the additional integral of motion is  elementary, while    for  
marble rolling it is not  elementary. }
I call by  ``Nose" function their expression  $N(\theta) = (I_1\, \cos^2 \theta + I_3\, \sin^2 \theta + m \, z_C^2(\theta))^{1/2}   $    where  $\theta$ is the nutation and  $z_C(\theta)$ is the    center of mass height. 
 $N(\theta)$ appears  somewhat miraculously in  the process of  the almost symplectic reduction. I work   in   space frame    using  the Euler angles $ \phi \, \text{(yaw)}, \, \theta \, $. 
The reduction to 1 DoF  is  done in  two stages:   first,  reduction by  the group  $SE(2) = \{ (x, y, \, \phi) \} $ to the Poisson sphere $T^* S^2 = \{ (\theta,   \, \psi) \}$ with 
   almost symplectic 2-form   $\Omega_{NH} = dp_\theta \, \wedge  d\theta + dp_\psi \, \wedge d\psi + J \cdot K$. The semi-basic term  is  $J \cdot K =  - p_\psi \, (d \log(N(\theta))  \wedge d\psi$.   
It follows that   $\Omega_{NH} $  is conformally symplectic in the sense that $d(\frac{1}{N}\, \Omega_{NH} ) = 0. $
The conserved quantity  due to   the $S^1$ symmetry about the body axis is $\ell = N(\theta)\,\sin^2 \theta\,  \dot{\psi}$, yielding  the  desired  reduction to    $(\theta, p_\theta)$.  Further simplification results by taking the  new  time $dt = \sqrt{B(\theta)} \, d\tau, \,\, \text{with}\,\, B =  I_1 + m \, |CP|^2 $  where  $P = (x,y)$ is  the point of contact. 
  One gets  finally 
  $H = \frac{1}{2} \tilde{p}^2_\theta  + V(\theta), \,    V(\theta) = \ell^2/2 \sin^2 \theta      + m g\, z_C(\theta)\, \,\text{with}\,\,
  \tilde{p}_\theta  =  p_\theta/\sqrt{B} $ and usual symplectic form $ d\tilde{p}_\theta\wedge d\theta$.    The moments of inertia $I_1, I_3$  reappear in   the reconstruction. 
Some  initial observations are  presented for the torus.   Full reconstruction is in order.

\vspace{2mm}

\noindent AMS MSC(2020) ; 37J60, 70F25, 58A15, 58A30 .

 }

\newpage
 
 {\small 
 \tableofcontents
 }

 \newpage

  \section{Introduction} \label{introduction}
  
$\,\,\,\,\,$     
 In choosing this theme for  this submission  I   wanted   to honor the living legacy of Alexey Borisov,   by revisiting with a different method (the \textit{almost symplectic reduction   for  
   Chaplygin systems})   a  very nice result  appearing (almost in passing)  in  the fundamental work  \cite{Borisov2008}, section 5  by  A. V. Borisov,   I. S. Mamaev and  also with    I. A. Bizyaev  in  \cite{Borisov2013}, section 3.1,   about \textit{rubber rolling of solids of revolution over the plane and the sphere}.  
 I guess the first ``rubber paper"  by Alexey was   \cite{BorisovMamaev2007}, in 2007, with I. S. Mamaev.     
 Solids of revolution  were suggested in  \cite{EhlersKoiller2007}, section 4.4. 
 
  No new results  appear here, however  relations with current studies are given 
 in the appendix (by Luis Garc\'ia-Naranjo) 
and the final section points some  research directions (besides a few  queries along the text). I I  present an extended  bibliography. I hope that the the almost symplectic reduction %
could be   useful  also for higher dimensions, see   \cite{Jovanovic2018}, \cite{Naranjo2019}.

 The example of the ellipsoid  of revolution over the plane was settled  by  Alexander Kilin and Elena Pivovarova  \cite{KilinPivovarova2024}. 
 I asked  them  about   the status  for the torus,
 since they had  already  done  similar problems in \cite{KilinPivovarova2017, KilinPivovarova2018,KilinPivovarova2019}. 
They told me  that  the torus  was not considered  in   detail yet,    
but they have   preliminary results   about   bifurcation diagrams.    I look forward for their study.

The no-slip  constraint for the torus appears in the very beginning of  the treatise by Neimark-Fufaev   \cite{NeimarkFufaev}, Chapter II-2. The marble case  was recently analyzed from
    the Engineering perspective in \cite{Chica2014}, \cite{Agundez2023}.      Actually  it is  easier to work out the   general surface of revolution and then particularize for the rubber torus - certainly one of the simplest  examples. The simplest is the spherical body with an axisymmetric distribution of mass, the ``rubber Routh sphere".  ]). 
The geometry of the first Chaplygin
reduction (or compression) for this example is described in section 4.1 of  \cite{Naranjo2019}.
\smallskip \smallskip

\noindent {\it  Historical note.}  According to \cite{KilinPivovarova2024} the rubber rolling appeared already in two notes by Hadamard in the end of   a book by  Appell \cite{Appell}\footnote{\url{https://gallica.bnf.fr/view3if/ga/ark:/12148/bpt6k82007h/f2}} , see also
 Hadamard, 1895  \cite{hadamard1895mouvements}. The theme  was developed in Beghin, 1929 \cite{Beghin1929}. There  appears  the result  that  in rubber rolling corresponding curves have equal  geodesic curvatures
 There are many ancient and modern references in   \cite{Chica2014} and \cite{OReilly1996}.

  \subsection{Rubber rolling  by  the RCD  school } \label{kilin}

    Kilin gently summarized for me their perspective. Following a tradition going back to Chaplygin, the equations and reduction are done  in Lagrangian
representation, with  redundant variables using the body frame. 
\vspace{2mm}
\newpage

One directly searches   a  Hamiltonian/Poisson structure, 
or at least  conformally Hamiltonian  (usually after a reduction). 
Here, to be conformally Hamiltonian    means here that there is an almost symplectic or  almost Poisson structure,  that becomes  symplectic (or Poisson) when multiplied by a function of a reduced coordinate, which is equivalent to a time change.

 See for instance \cite{Garcia-Naranjo2024}.  This notion of ``conformality"  contrasts with \cite{Calleja2013}. 

  The configuration space for rolling over the plane is
$ Q =\{(P, e_1, e_2, e_3)\}= \R^2 \times SO(3)  $ where $P$ is the contact point in the plane.
One can find reduced equations  for $(\gamma, \dot{\gamma}) \in TS^2 $  where $\gamma$ is the unit vector $e_z$ in the space frame seen in the body frame.   
$\gamma $  is usually   called   the  \textit{Poisson vector}.

To achieve this reduction, one takes quasi-velocities  $v$ and $\Omega$ - the velocity of the
center of mass
and the angular velocity, respectively, referred to the body. 
 Using the method of undetermined multipliers, one writes the
(Lagrangian) equations
of motion on $TM$  using  these quasi-velocities. 

By construction, these
equations preserve the
constraint equations. Also, they are invariant under the action of $SE(2)$.
More precisely,  $v$ and $\Omega$ are obviously invariant under $SE(2)$, so the
corresponding reduction is trivial,  and moreover 
 the equations for $v, \Omega$
and $\gamma $  decouple. In more detail, it goes as follows:

\begin{enumerate}[i)]

\item  Thanks to the $SE(2$)-symmetry, the equations for $\Omega$ and $\gamma$ are decoupled from the equation
for $v$ (which expresses the no-slip constraint). 
\item  Using the no-slip constraint, one drops the equation for v obtaining a system for $(\Omega,\gamma) \in
\R^3 \times \R^3$. The evolution equation for $\gamma$ is kinematical (Poisson equation  $\dot{\gamma}= \gamma \times \Omega$)
whereas the evolution equation for $\Omega$ is dynamical (Newton’s law). These equations have
the geometric integral $F_0 = ||\gamma||^2  $ and also, by construction (appropriately considering
D’Alembert’s principle), the ‘rubber constraint’ integral $F_1 = \gamma \cdot \Omega$. The restriction of the
equations to the physical invariant set $\Sigma = \{(\Omega,\gamma) \in \R^3 \times \R^3 : F_0 = 1, F_1 = 0 \}$ defines
the $SE(2)$-reduced equations of motion. It is clear that $\Sigma$ is diﬀeomorphic to $TS^2$. 
\item  Using the rubber constraint $\gamma \cdot \Omega = 0$, one may write $\Omega =  \dot{\gamma} \times \gamma$ along $\Sigma$ (take cross product
with $\gamma$ on both sides of the Poisson equation and use standard vector identities). This
allows one to write the equations on $\Sigma$ as a second order diﬀerential equation for $\gamma \in S^2$ or,
equivalently a (second order) vector field on $TS^2$. (This step is also performed by Borisov
and collaborators, see section 2.2 in [2]). This is precisely the compression of the Chaplygin
system in the velocity, or Lagrangian, perspective. 
\item For bodies of revolution there is an additional $S^1$ symmetry of the vector field on $TS^2$
corresponding to the action
$  (\gamma, \dot{\gamma}
  \rightarrow (R\gamma, R \dot\gamma), $ 
  or
  $  (\gamma,\Omega) \rightarrow(R\gamma,R\Omega)
  $ 
  if one wants to work in redundant variables 
where 
$$R= \left( \begin{array}{lll} \cos \alpha & - \sin \alpha  & 0 \\ \sin \alpha & \cos \alpha & 0 \\
0 & 0 & 1 
\end{array} \right).
$$
Independent invariants of the action are $\gamma_3, \dot{\gamma}_3 $ and  $\dot{\gamma} \cdot (\gamma \times e_3)$  (or $\gamma_3, \Omega_3, \Omega \cdot (\gamma \times e_3)).$
Considering that $  \gamma_3 = \cos \theta$, and there is an additional first integral, this implies that the
ultimately reduced system is a 1 DoF system for $\theta$ (this is not entirely obvious and some
work needs to be done to clarify this, which is an important point of the paper).

\end{enumerate}

This approach is genuinely geometric, following a long historical tradition  carefully annotated
 in the papers by Borisov, and his colleagues and students.   
Some rubber rolling papers by Alexey Borisov and  his associates:  
 \cite{Borisov2013},    \cite{BorisovMamaev2007},  \cite{KilinPivovarova2024},  \cite{KilinPivovarova2017},  \cite{KilinPivovarova2018},    \cite{Borisovetal2013},  \cite{Bolsinov2012},  \cite{Borisov2016}, \cite{Mamaev2020}, \cite{Moskvin2010}.
 
 \vspace{2mm}

Kilin wrote to us, in conclusion:

 ``Both in the classical (marble) nonholonomic model and in the rubber model, this
method works well. In the classical model,
the reduced system admits two integrals linear in velocities, whereas in
the rubber model it admits an integral and one no-spin constraint.
In both cases, after reduction  
 we obtain a system with one degree of freedom.  

Remarkably, the inertia $I_3$ only appears in the reconstruction of the  rotation angle $\psi$ around the body axis $e_3$,  a fact which is
valid for all bodies of revolution \cite{KilinPivovarova2024}. 
 I suppose the first time the reduced equations of motion (with 1 DoF) for arbitrary body of revolution  where derived in  \cite{Borisov2008}\footnote{One can find also in \cite{Borisov2013}.}.

There is no parametric analysis of the reduced system in this paper. But from Eq (5.2) [in that paper]  it is clear that this reduced system does not depend on $I_3$. We noticed this fact explicitly for problems of motion of truncated ball and ellipsoid of revolution in our papers  \cite{KilinPivovarova2024}, \cite{KilinPivovarova2017}, \cite{KilinPivovarova2018}. 

But we did not formulate this fact as some general theorem.  Actually there is even more interesting fact: 
The bifurcation diagrams  for an  arbitrary body of revolution does not depend on both $ I_1 $ and $I_3.$
This  follows from the independence on $ I_1 $ of the reduced system after some changing of time and $  p_\theta$. In our case new time and $p_\theta$  are defined by
$$ \frac{d t}{d \tau}=\sqrt{B} \,\,\,\,, \,\, \,\,  p_\theta=\sqrt{B} \, \dot{\theta} $$
Again we did not formulate the general theorem but have noticed it for  all considered systems.'' 
\vspace{2mm}

  \subsection{Alexey Borisov, RCD  and Brazil }    
$\,\,\,$The scientific achievements of Alexey Borisov,    
  described in \cite{Borisovinmemmory},  show  how  untimely was his  loss.    I met  Alexey personally   in 2006, by his very kind invitation to participate, together with Kurt Ehlers,  in the IUTAM Symposium 
on Hamiltonian Dynamics, Vortex Structures, Turbulence
held in Moscow, 25-30 August, 2006 \cite{Iutam2006}.  We presented a discussion on 2-3-5 distributions and the  related sphere-sphere rubber rolling problem.  We were thrilled to meet so many congenial people.

   \begin{figure}[h] \label{fig:Borisov1}
\hspace{-1.5cm}
            \includegraphics[width=1.1\textwidth]{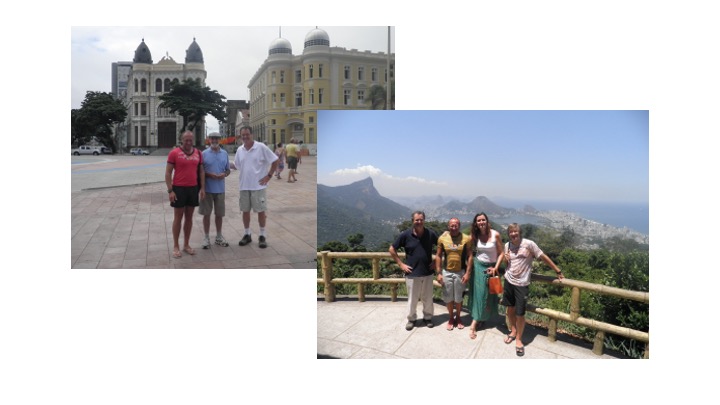}  \label{fig:RecifeRio}
\vspace{-20mm}
 \caption{\small {Alexey Borisov and Ivan Mamaev in  Rio, hosted by Stefanella Boatto  and Jair and  a few days later in  Recife, hosted by Hildeberto Cabral, 2011. }}
 \end{figure}

I could  reciprocate the hospitality only  five years later.  After a  billiards conference in S\~ao Paulo, Borisov and Mamaev  came to Rio and Recife
(Fig.1 \ref{fig:Borisov1}) to give talks and meet researchers from IMPA and the Federal Universities. 
In Recife  Alexey participated in the committee for the Ph.D. thesis defense of Adriano Regis.
 Together with two Brazilian colleagues,  Jorge Zubelli and Hermes Gad\^elha, I went to  Izhevsk  in 2012, for  the IUTAM Symposium ``From Mechanical to Biological Systems: an Integrated Approach.    
 Mathematics and Music - guitar and accordion  and  folk dance   were wonderful.  In  2018 I had again    a great time in Moscow with my wife Rosa. 
   Alexey  participated in  the scientific committee of the online seminars  that the Recife group  started in 2020. RCD editorial board has encouraged   young Brazilian students from the Recife group to publish their work, recently involving the Krein-Gelfand-Lidskii theory of auto-parametric resonance.\\

 \subsection{A  very nice result  from \cite{Borisov2008} and   \cite{Borisov2013} }  
    
\begin{figure}[h] \label{fig:figdata}
\centering
\includegraphics[width=0.4\textwidth]{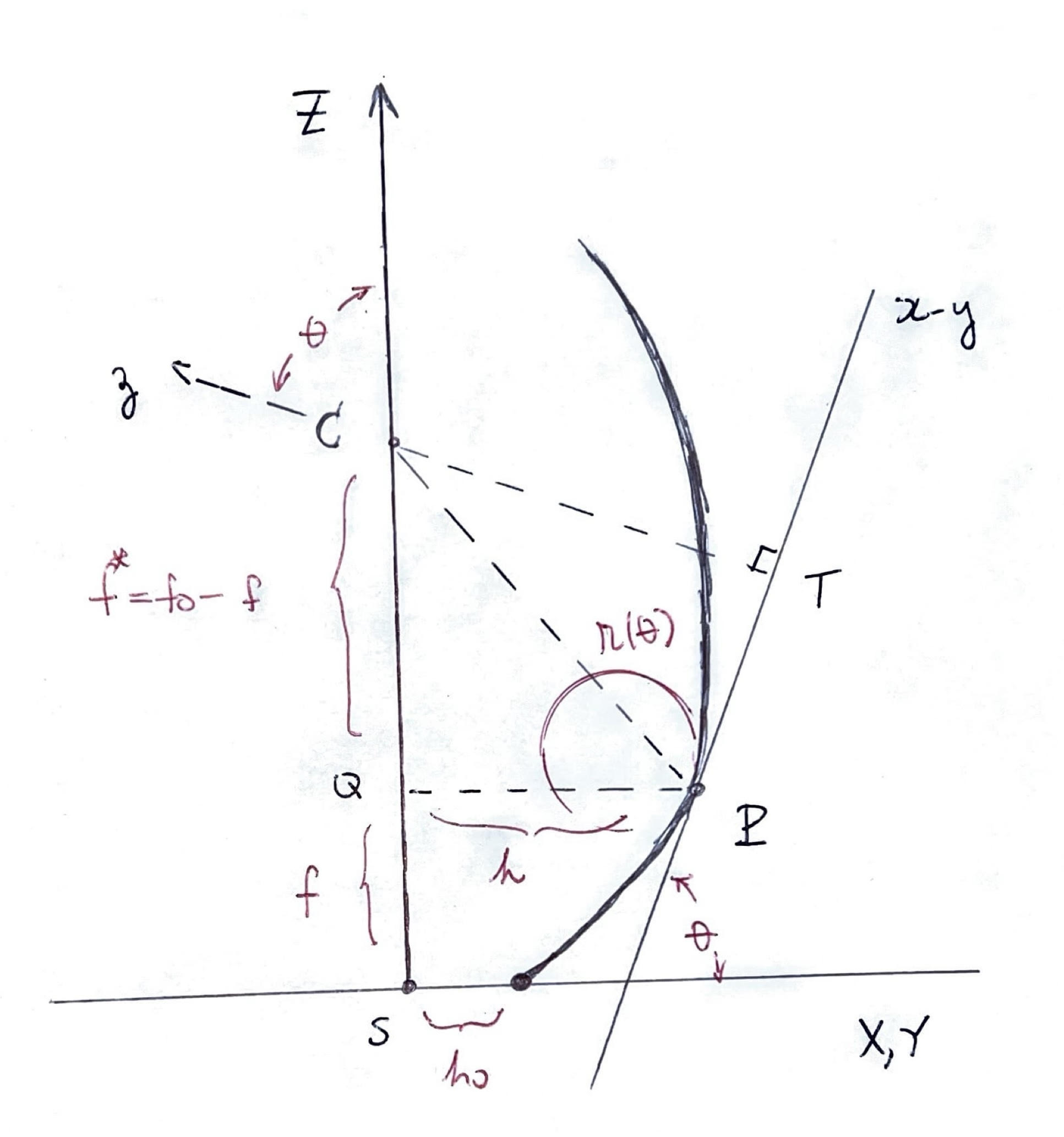}
\vspace{-5mm}
\caption{{\small The standard position.}}
\end{figure}

   \begin{teon}   Let   $\theta$  be the angle between  the moving body axis $e_3$ with the vertical  axis $e_z$ (nutation).  The  Hamiltonian for the  reduced  1 DoF  in  $(p_\theta, \theta)$,  with the usual symplectic structure $dp_\theta \wedge d\theta$  is
\begin{equation}
\tilde{H} = \frac{1}{2} \left( \frac{p^2_\theta}{B(\theta)} + \frac{\ell^2}{\sin^2 \theta}  \right) + m g\, z_C(\theta)\,\,\,,\,\,\, B =  I_1 + m \, |CP|^2(\theta) . 
\end{equation}
\centerline{($z_C$ the height  of the center of mass and $ |CP|$ its distance   to the contact point  $P$).}
Reconstruction is done using the conserved  $\ell = N(\theta)\, \sin^2 \theta\,   \dot{\psi}\,\,, \, N = [I_1\, \cos^2 \theta + I_3\, \sin^2 \theta + m \, z_C^2(\theta]^{1/2} $  (where $\psi$ is the precession angle),  
and the nonholonomic constraints: no-twist for 
the proper rotation $\phi$  and  no-slip for the contact point in the plane  (see Fig.2 \ref{fig:figdata})

The  change of time  $dt = \sqrt{B(\theta)} \, d\tau$, pointed out to me  by A. Kilin,   leads to the  traditional Newton's equations  (where  $' = d/d\tau$) 
\begin{equation}
\theta'' = -  dV/d\theta, \,\,\,  V  = \frac{\ell^2}{\sin^2 \theta}    + m g\, z_C(\theta) . 
\end{equation}
\end{teon}

\vspace{2mm}

 In re-deriving this result, instead of  using the body  frame, more usual in such problems,  I  worked  directly in the  space frame\footnote{We can't refrain from using a  boxing metaphor:  ``a fighter never falls.  It is the canvas  that tilts   and goes up''.}.
In the torus example I  stopped far from f the  final goal,  which is  to  describe  all  the  possible qualitative regimes, with bifurcation diagrams as  the parameters vary,   pictures of solutions,      Poincar\'e sections, etc.,  and  also  obtain the solutions analytically.

\textit{Such  thorough analysis is  always present  in  the  papers by  Alexey Borisov with  his partners and students.
 We hope this  study will be done  
for  the torus soon by Kilin and Pivovarova. } 

\vspace{2mm}

\subsection*{Data used in the sequel}

\begin{itemize}
 \item
 Geometry:  the meridian is defined by  the radius of curvature  $r(\theta) \geq 0, \,  \theta \in [0,\, \pi]$,  where the nutation angle  $\theta$  can be also seen as the angle of the tangent with the horizontal plane, and a parameter  $h_o \geq  0$, the radius of the parallel at $\theta = 0$.  For the torus, $h_o = R$ and $r(\theta) \equiv r$. 
\item  
 Dynamic quantities: the height  $f_o$ of the center of mass in the standard position (for the torus, $f_o = r $);  the 
 mass $m$ and inertias $I_1= I_2 < I_3 < 2I_1$. 
 \end{itemize}

\newpage

\vspace{- 3mm}

It is interesting  that from  the functions $|z_C(\theta)|$ and $|CP|(\theta)$  one can recover the meridian profile and the position of the center of mass. It is  like a `compass and ruler' construction:  take $C$ for center.  For each  angle $\theta$ draw the perpendicular segment $CT$  of length $z_C(\theta)$  (so $T$ is the projection to the plane viewed in the body frame) and then, along the $\theta$-ray , draw the segment $TP$,  whose length we also know: 
  $|TP|^2 = |CP|^2 - |CT|^2$.

\bigskip

\section{Set up  with Euler angles  in the space frame }  \label{setup} 

The configuration space can be thought of in two ways. Either as 
\begin{equation}  Q=\{( e_1, e_2, e_3 \,,P )\}= SO(3) \times \R^2,
\end{equation}
 with  $P=(x,y)$  the contact point in the plane and $e_i, i=1,2,3$ the body  principal axis of inertia, or  alternatively 
 \begin{equation} \label{Q1}
Q = \Sigma \times SE(2) =  \{ (u, v ;  x , y, \phi)          \}  .
\end{equation}   
where   $(u,v)$  parametrizes  the corresponding contact point of the body surface,  
and $\phi \in S^1$  describes the relative rotation of the tangent spaces about the common normal.  

\begin{figure}[t] \label{fig:EuleranglesArnold}
\centering
   \includegraphics[width=0.45\textwidth]{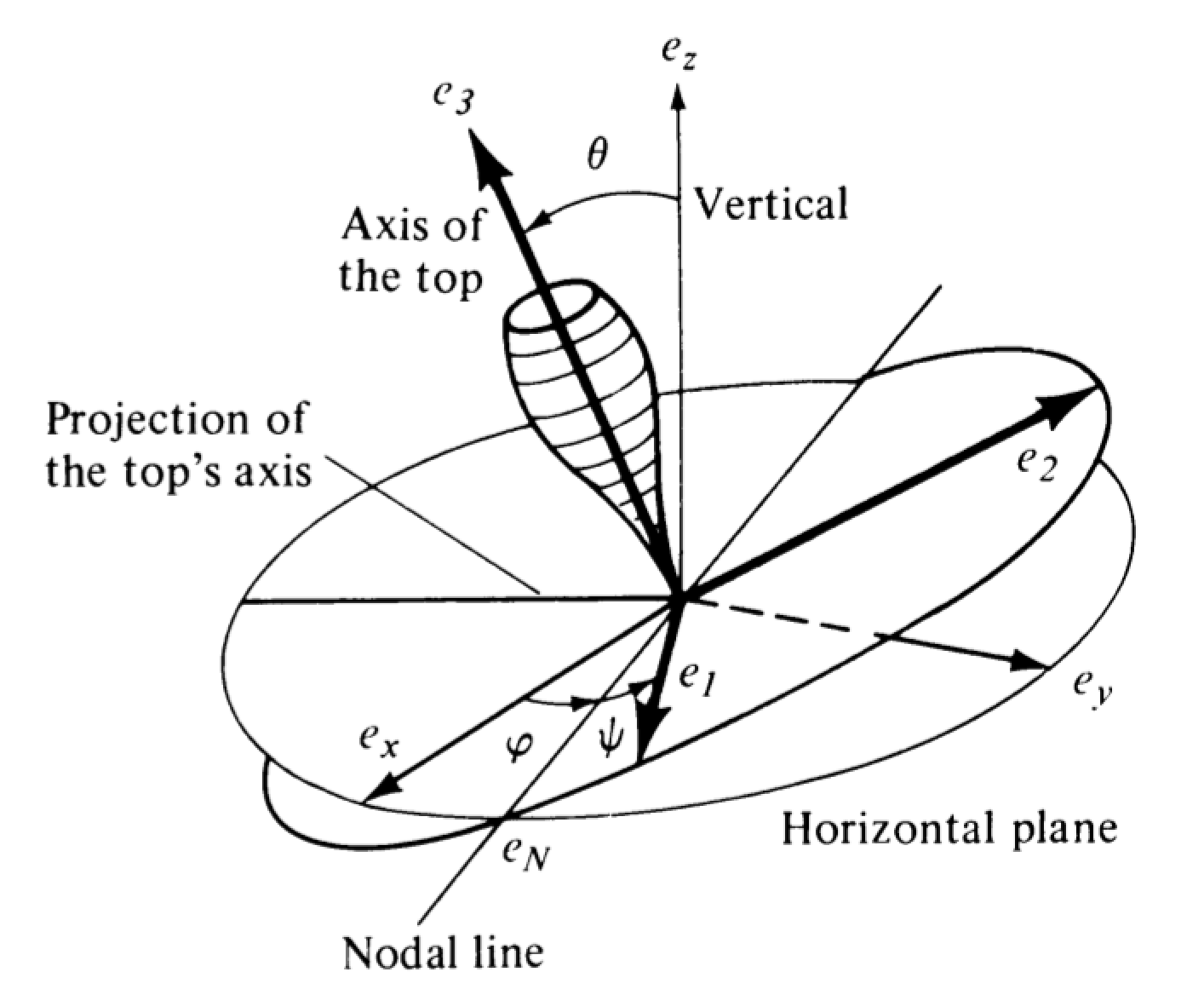} 
\vspace{-1mm}
 \caption{\small{Euler angles for Lagrange's top. Taken from  Arnold's \cite{Arnoldbook} (Fig. 126 pg. 149).     The angle $\phi$ from $e_x$ to  $e_N = (\cos \phi, \sin \phi, 0) \equiv e^{i \phi}$ (a choice in    the nodal line)    produces an isometry  between the horizontal  
  plane    and plane $e_1-e_2$. 
A choice between $\phi$ and $\phi + \pi$  is needed.  Note that  $e_N^\perp = + i \, e_N $   is  always in the opposite  direction to the projection of $e_3$.}}
 \end{figure}

Let $\phi$ the angle from the $x$-axis  to the nodal line, which is the intersection of the plane $e_1-e_2$ with the $x-y$ plane,  Fig. 3 \ref{fig:EuleranglesArnold}.    The three coordinates $(x, y , \phi)$ form  an element of $SE(2)$.   

 As   parameters $(u,v)$ I take the other two Euler angles $(\psi, \theta)$ of the attitude matrix of the  body.  For that  one   uses    the Gauss map $ \Sigma \rightarrow S^2$, as follows:  
 
\vspace{5mm} 
\noindent  {\bf Using the  two other Euler angles $\theta , \psi$  to parametrize the surface.} 

The standard position for the body surface is  $\Sigma_{st} $  is when   
   the  body axis  $e_1, e_2, e_3$ are aligned with fixed  coordinate axis $e_x , e_y, e_z$.  
One may place the contact point at the origin of the $x-y$ plane.
I  assume the surface to be   convex (Gaussian curvature $\geq 0$), so the  Gauss mapping $${\rm Gauss}: (u, v) \in \Sigma_{st}  \rightarrow  (\psi, \theta) \in S^2$$ is invertible in the range of interest. 
\begin{dfn}  
$\,\,\,\,\,\, \gamma = {\rm Gauss}(u,v) \,\,\,\, \, \text{{\rm is called the Poisson vector}}.$
\end{dfn}

\subsection{The  ZXZ Euler angles.}  I  take  the  basic rotation matrices in the order $ Z_\psi X_\theta Z_\phi$  (see  e.g. Arnold's Mathematical Methods of Classical Mechanics \cite{Arnoldbook}, section 30)\footnote{Beware that in papers of the Izhevsk group  the roles of $\phi$ and $\psi$ are interchanged.}.  To carry the stationary $(e_x, e_y, e_z)$  
   into the moving frame  $(e_1, e_2, e_3),$  one does  
  three rotations: 
 \begin{itemize}
\item  angle $\phi$ around the $e_z $ axis. $e_z$ remains fixed, and $e_x$ goes to $e_N= (\cos \phi, \sin \phi, 0)$. 
\item  angle $\theta$  around the $e_N$  axis. Under this rotation, $e_z$  goes to $e_3$, and $e_N$  remains fixed. 
\item     angle $\psi$  around the $e_3$ axis. Under this rotation, $e_N$ goes to $e_1$ , and $e_3$ stays fixed. 
 \end{itemize}
 
 \vspace{2mm}
 
\noindent Since  $e_x  \to e_1$,   $e_z  \to e_3$,  $e_y$ goes to $e_2. $
They form the columns of the attitude matrix
\vspace{2mm}
\begin{equation} \label{Rmatrix}
\begin{bmatrix}
\cos\phi \cos\psi -  \cos\theta \sin\psi \ \sin\phi &
-\cos\phi  \sin\psi - \sin\phi \cos\theta \cos\psi &
\sin\phi \sin\theta \\
\sin\phi \cos\psi + \cos\phi \cos\theta \sin\psi &
- \sin \phi \sin\psi + \cos\phi \cos\theta \cos\phi &
-\cos\phi \sin\theta \\
\sin\theta \sin\psi &
\sin\theta \cos\psi &
\cos\theta
\end{bmatrix}
\end{equation}
The angular  velocity vector in space  is
\begin{equation}
\begin{split}
&\mathbf{\omega } = \dot{\phi} \, e_z + \dot{\theta} \, e_N + \dot{\psi} \, e_3  =  \omega_1\, e_x + \omega_2\, e_y  + \omega_3\, e_z \\
 \end{split}
\end{equation}
\begin{equation}
\begin{split}
&  \omega_1 =  \cos \phi \, \dot{\theta} + \sin \phi\, \sin \theta\, \dot{\psi} \,\,, \,\,
 \omega_2 =  \sin \phi \, \dot{\theta}  - \cos \phi \, \sin \theta \dot{\psi}  \,\,, \,\,
  \omega_3 =  \dot{\phi} + \cos \theta \, \dot{\psi}
 \end{split}
\end{equation}

\vspace{2mm}

\subsection{The no-twist constraint}

\begin{prp}The  no-twist condition  $\omega_3 =  \omega \cdot e_z = 0$   is  
\begin{equation}
\dot{\phi} = - \cos \theta \, \dot{\psi} 
\end{equation}
\end{prp}

\begin{rem}  In this paper I  will not  need  the body representation
$\mathbf{\omega} = \Omega_1\, e_1 + \Omega_2  \, e_2 + \Omega_3 \, e_3\,\, $
\begin{equation} \label{Omegaformula}
  \Omega_1 =  \dot{\phi} \sin \theta \sin \psi + \dot{\theta} \cos \phi \,\,, \,\,
 \Omega_2 =  \dot{\phi} \sin \theta \cos \phi - \dot{\theta} \sin \psi \,\,, \,\,
  \Omega_3 =  \dot{\phi} \cos \theta + \dot{\psi}
\end{equation}
except for the observation that, under the no-twist condition, 
\begin{equation} \label{ufa}
  \Omega_3 =  \dot{\phi} \cos \theta + \dot{\psi} =  - \dot{\psi} \cos^2 \theta + \dot{\psi} =  \sin^2 \theta\, \dot{\psi} .
\end{equation}
I will use this observation in  (\ref{ufaufa}).
\end{rem}

\subsection*{The no-twist constraint resembles  the  Coriolis effect } 

This is a purely  kinematic  onbservation.  Suppose the contact point on the surface  describes a  parallel in the surface corresponding to  radius $h(\theta)$. The well known formula for    geodesic curvature $ \kappa_g =  (\gamma'(t) \times \gamma''(t)) \cdot n/|\gamma'(t)|^3 $
gives 
\begin{equation}
 \kappa_{\theta} = \frac{\cos \theta}{h(\theta)} . 
\end{equation}  
 
In rubber rolling  the corresponding curves have the same geodesic curvature, so the contact point  travels a circle in the plane. Due  to no-slip, the scalar velocities of the contact point in both surfaces are the same.   The radius of the planar circle is
$  h(\theta)/\cos \theta . $  Thus the projection of the center of mass describes another circle, but  with the same center.  

 The ratio between the parallel in the surface and the planar circle is  $ 1/\cos \theta$. Thus the  radius of the circle in the plane is   bigger than the radius of the parallel.   
 Motorcycle racing is hard! 
 
As if to compensate, corresponding to  the angular velocity $\dot{\psi} $  (say $> 0$) of the wheel, the tangent vector to the circle in  the plane   rotates \textit{counterclockwise}, in with angular velocity     $$\dot{\phi} = - \cos \theta\, \, \dot{\psi}. $$

 \subsection{The no-slip  constraint} 
 
 Here I consider only   surfaces of revolution,  for which $\psi$   is  ignorable.  Let $r(\theta)$ be the local radius of curvature, and $h(\theta)$ the distance of the contact point  to the symmetry axis.

 \begin{prp}
 \begin{equation} \label{noslip}
  (\dot{x}, \dot{y}) = - h(\theta) \dot{\psi} \, e_N  -  r(\theta) \dot{\theta} \, e_N^\perp .
  \end{equation} 
  \end{prp}

\subsection{Inverting the Gauss map for surfaces of revolution} 

 In  a  standard position  I  take   $e_3 = e_z$ pointing up.  One has  the relations
 \begin{equation} \label{miracle}
 r(\theta) \cos \theta = dh/d\theta \,\,,\,\,\,  r(\theta) \sin \theta = df/d\theta
 \end{equation}
 for the meridian    given by the pair of  $(h(\theta) ,\, f(\theta))$ with  $\, f\,$  increasing along the axis   in the direction of $+e_3$. 
 The proof is elementary: 
   $ds^2 = df^2 + dh^2\,,\,\, ds \sin \theta = df\,, ds \cos \theta = dh. $  Now just recall $r(\theta) = ds/d\theta. $

\begin{figure}[h] \label{fig:basicpict}
\centering

    \includegraphics[width=0.55\textwidth]{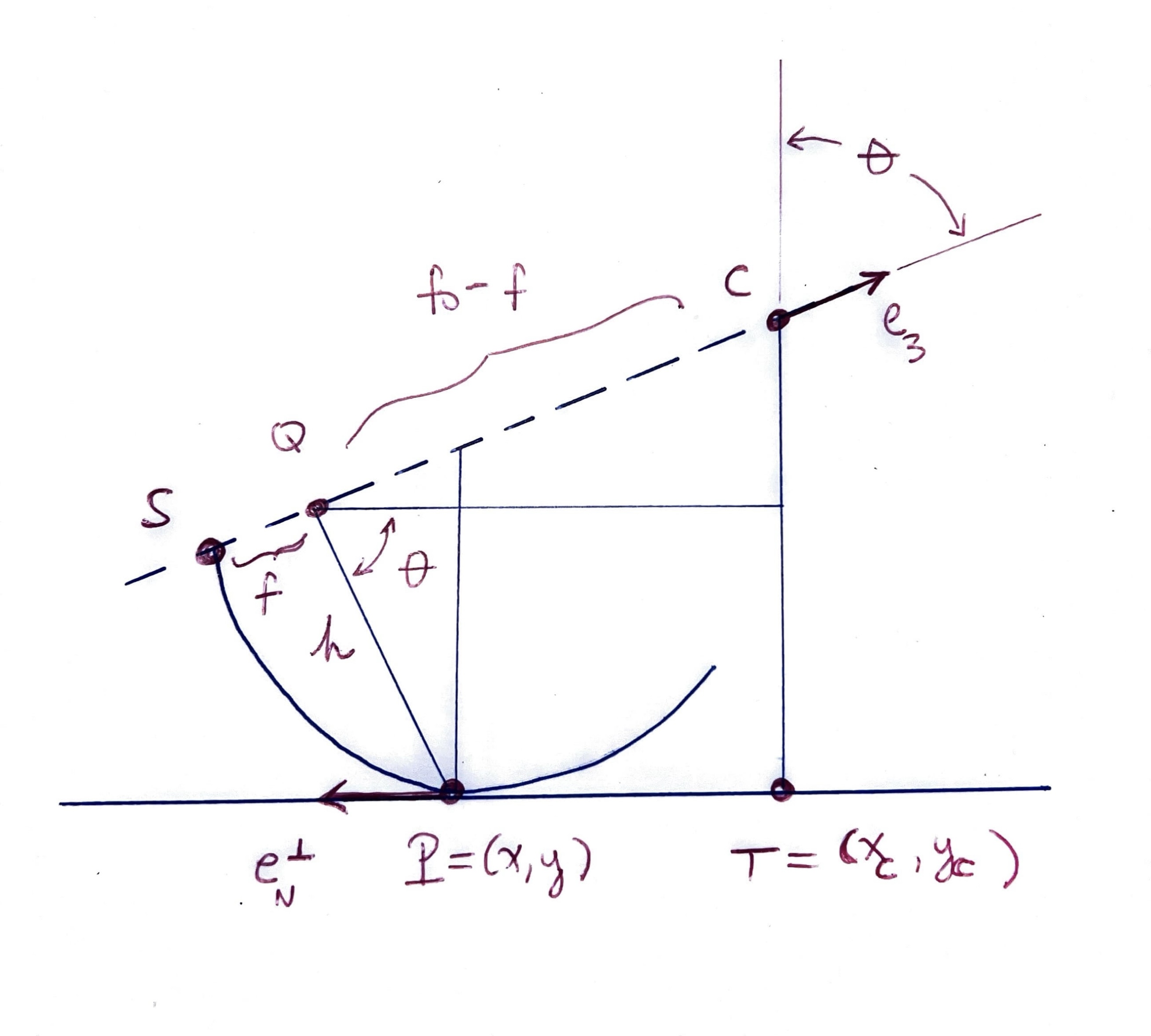} 

 \vspace{-1.1 cm}
 \caption{\small{Space frame viewpoint (compare with Fig. 2 \ref{fig:figdata}).   The vector  $e_N = (\cos \phi, \sin \phi, 0) $ is sticking out  of the page  and $e_N^\perp =  (-\sin \phi, \cos \phi)$ points in the direction \textit{opposite} of the projection of $e_3$ (see   (\ref{Rmatrix}).    Then  $z_C(\theta) = h(\theta)\, \sin \theta + f^*(\theta)  \, \cos \theta,\,\,  f^*(\theta) = f_o - f(\theta)$ and   $PT = \Lambda(\theta)  \, e_N^{\perp} $ with  $ \Lambda(\theta) =  h(\theta)\, \cos  \theta  -  f^*(\theta) \, \sin \theta. $   A bit surprisingly  $ dz_C/d\theta  = \Lambda(\theta)$.  Spoiler:  this will be important for   the ``miracle".    }}
  \end{figure}

One  can  construct the meridian with single integrations. At the south pole $(0,0,0)$, the meridian may start  at  $(h_o, 0 , 0)$ for some $h_o>0$. Then  (see 
 Fig.4 \ref{fig:basicpict})
\begin{equation}  \label{data}
h(\theta) = h_o + \int_0^\theta \, r(\theta) \, \cos \theta \, d\theta\,\,,\,\, f(\theta) =   \int_0^\theta \,r(\theta) \, \sin \theta \, d\theta
\end{equation}
Let  $f_o$  be  the height of the center of mass  $C$ when the body is  in the standard position.  Then  the vector  from $Q$ (the  projection  of  the of the contact point $P$
on the  axis)   to $C$  is given by 
\begin{equation}   \overrightarrow{QC} =   f^*(\theta) \, e_3, \,\, f^*(\theta) = f_o - f(\theta)\,\,\,,\,\,  0  < f_o < \int_0^\pi \, r(\theta) \, \sin \theta \, d\theta  \,\,  (= f(\pi)) .
\end{equation}

\begin{dfn}
\begin{equation} \label{Lambdadfn}
\Lambda(\theta) =  h(\theta)\, \cos  \theta  -  f^*(\theta)\, \sin \theta .
\end{equation}
\end{dfn}
\begin{prp}
\begin{equation}
 \overrightarrow{PT} = \Lambda(\theta) \, e_N^\perp . 
\end{equation}
\end{prp}

 \vspace{-0.5cm} 
 
 \vspace{-4mm}
 
 \subsection{For   use in section \ref{rubberroll}}
 
\begin{prp} The
height  of the center of mass  is 
\begin{equation}  \label{zC}
\begin{split}
& z_C(\theta) = h(\theta)\, \sin \theta + f^*(\theta)  \, \cos \theta
\end{split}
\end{equation}
\end{prp}

\begin{prp} \label{dzC}  (this formula  will allow the ``miracle"; please    do not  look  (\ref{truemiracle}) yet.)
\begin{equation}
  \frac{dz_C}{d\theta}  = \Lambda(\theta)
  \end{equation}
\end{prp}
\noindent {\bf Proof. }Due  to (\ref{miracle}) the terms involving  $h' $ and $f'$ cancel when differentiating (\ref{zC}).

\begin{prp}  \label{PTprop}   We will need for future use (in section \ref{compressedleg}): \begin{equation} \label{dLambda}
\begin{split}
 &  \frac{d \Lambda}{d\theta} =  r(\theta)  -   f^*(\theta) \cos \theta -    h(\theta) \sin \theta  \\
& |CP|^2  =   h^2 + (f^*)^2  = \Lambda^2    +  \, (r(\theta) - d\Lambda/d\theta)^2   \\
& h - \Lambda \, \cos \theta =  z_c \, \sin \theta
\end{split}
  \end{equation}
  \end{prp}

\begin{figure}[t] \label{fig:fig5}
\centering
\begin{minipage}[c]{0.5\linewidth}
\includegraphics[width=\linewidth]{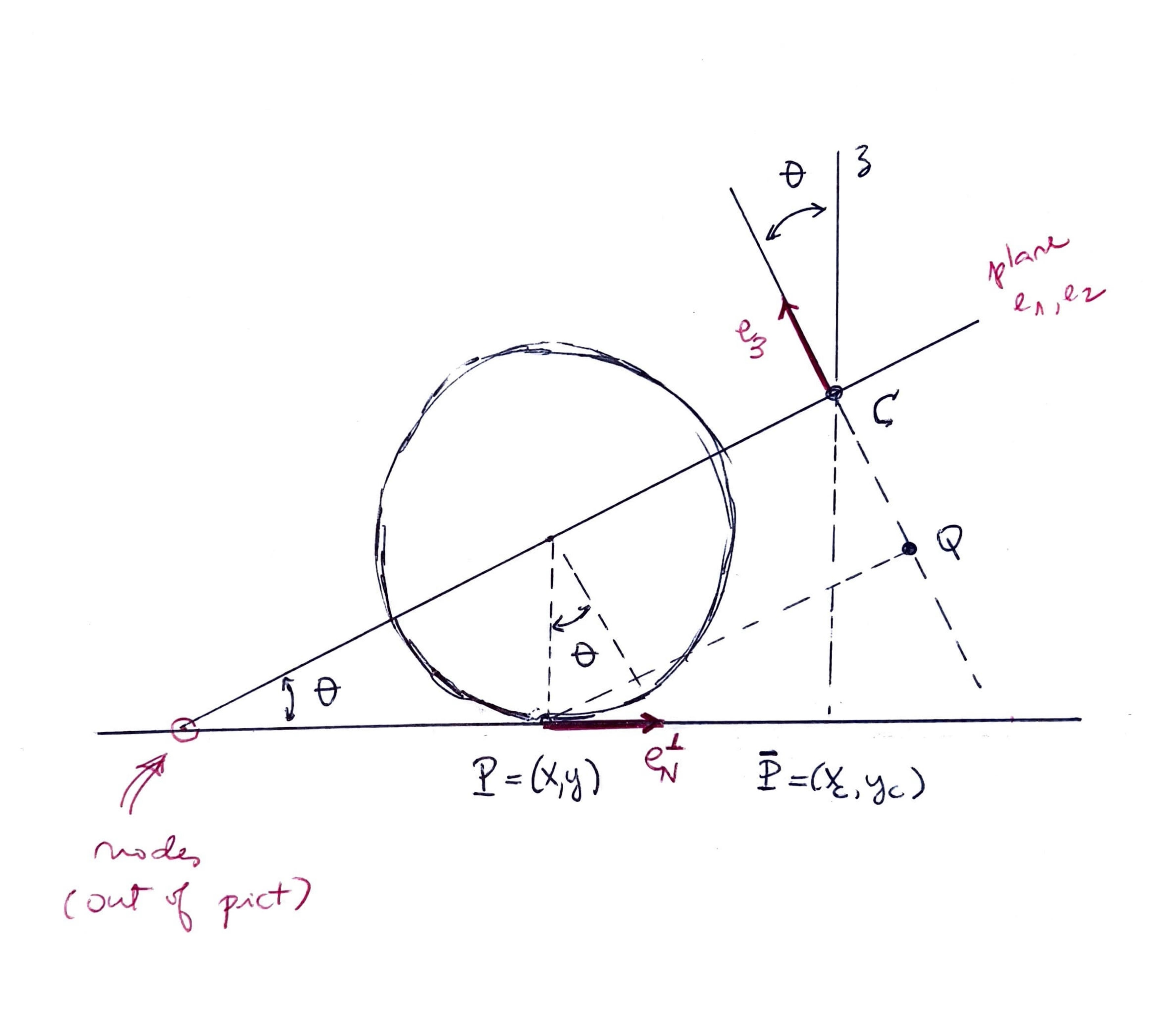}
\end{minipage}
\hspace{2cm}
\begin{minipage}[c]{0.3\linewidth}
\includegraphics[width=\linewidth]{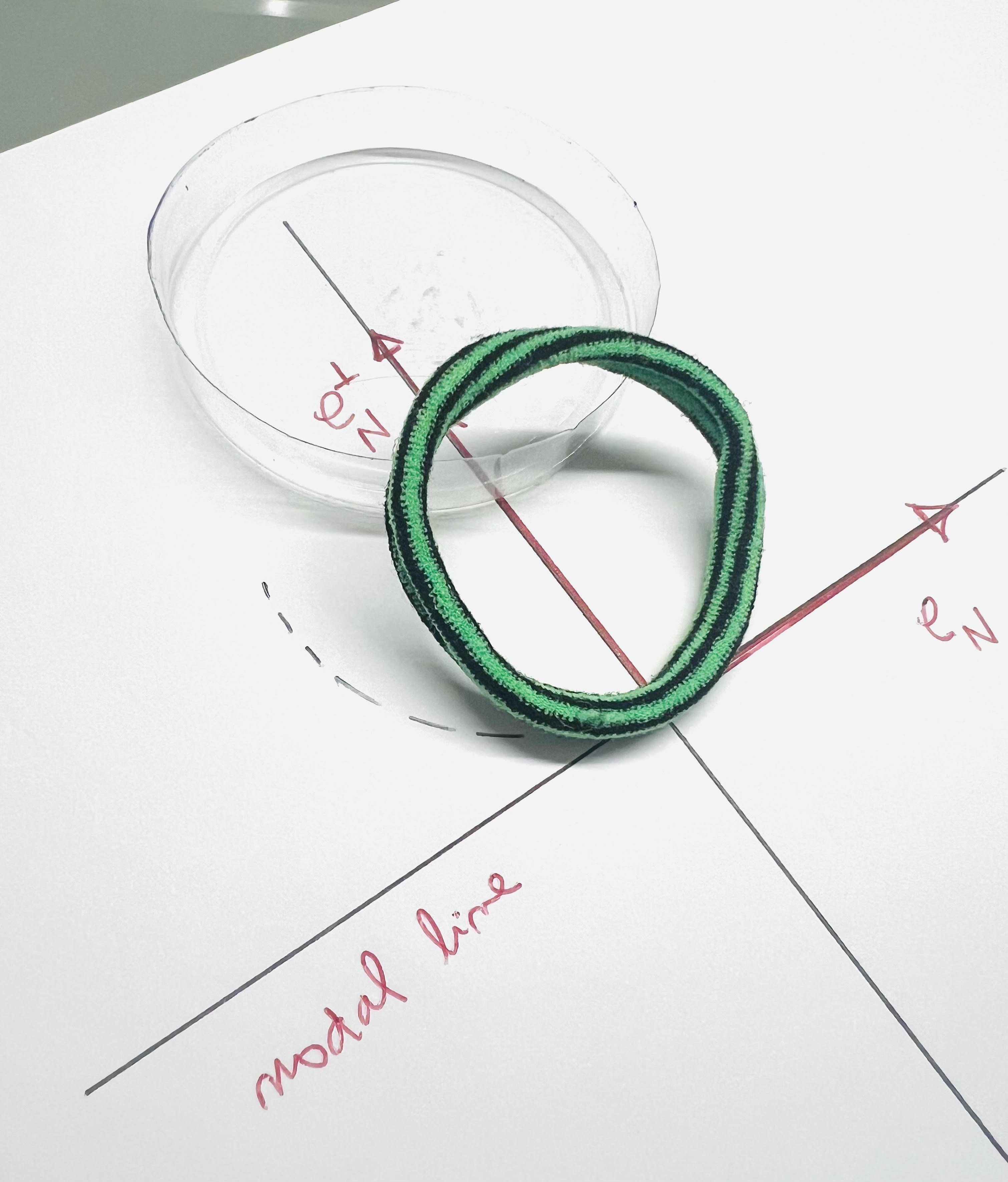}
\end{minipage} 
\caption{{\small The torus example.  In this picture the   symmetry axis  $e_3$  
 is  pointing to  the sky, so $ 0 < \theta < \pi/2$.  We keep the convention:   $e_N^\perp$ is opposite to the  projection of $e_3$. Left. Coordinates $\phi$ and $\psi$  are  ignored in this picture (they are ignorable).  
 The vector $e_N$  emerges out of the picture. Right.  When $\psi$  increases the torus rolls in the direction \textit{opposite} to $e_N$;  when  $\theta$  increases it rolls in the direction \textit{opposite } to  $e_N^\perp$. 
 }
\label{fig:torus-seen-from-eN}}
\end{figure}

\begin{rem} Notation in \cite{Borisov2013}, section 3.1.  In standard position, the surface of revolution is 
parametrized in terms of the \textit{ internal} normal vector $ \gamma$ as  
$ 
\bm{r(\gamma)} = (f_1(\gamma_3) \gamma_1, f_1(\gamma_3) \gamma_2, f_2(\gamma_3)) .
$ 
The correspondence is immediate: their vector $\gamma$ (which is the opposite of mine)
$   \gamma$ is the vector $- e_z$  seen in the body frame, thus it is the last row of  the attitude matrix (\ref{Rmatrix}):
\begin{equation}
  \gamma =  (\gamma_1, \gamma_2, \gamma_3) = - (\sin \theta \sin \psi,\, \sin \theta \cos \psi,\, \cos \theta) .
\end{equation}
The function $f_1(\gamma_3)$ in \cite{Borisov2013} is precisely our  $h(\theta) $ while  $f_2(\gamma_3)$    requires a small adjustment.  Note that the south pole corresponds to $\gamma = (0,0,1)$, i.e, $\theta=0$, so
\begin{equation}
f_2(\cos \theta) \,\, \text{is equal to our} \,\,\, f^*(\theta) .  
\end{equation}
\end{rem}

\begin{exan}  \label{torusexa}
For the torus with parameters $R>r$  we have $h_o = R, \, f_o = r, r(\theta) \equiv r$, hence
\begin{equation}
h(\theta) =  R + \int_0^\theta \, r \cos\theta \, d\theta = R + r \sin \theta \,\,,\,\, f(\theta) = r (1- \cos \theta) 
\end{equation}
and a quick calculation gives, as expected  
\begin{equation} z_C = (R+r\sin \theta) \sin \theta + (r - r(1-\cos \theta)) \cos \theta  =  r + R \sin \theta .
\end{equation} 
\end{exan}
As a check  (see Fig.5  \ref{fig:fig5})
\begin{equation} \label{Lambdatorus}
\Lambda(\theta)  = h(\theta)\, \cos  \theta  -  f^*(\theta)\, \sin \theta = (R + r \sin \theta) \, \cos \theta  -  r \, \cos \theta \, \sin \theta = R \cos \theta . 
\end{equation}

 \newpage

 \subsection{The `skidding' Lagrangian}  \label{skidding1}
 
 Recall that the data   are the functions $(h(\theta), f(\theta))$  given by (\ref{data}),  that can be constructed from the curvature function $\kappa(\theta) = 1/r(\theta)$ and the
parameters $h_o , \, f_o$.   From (\ref{zC}) we already have the potential energy
\begin{equation} \label{zC1} 
V(\theta) = m g\, z_C(\theta)\,\,, \,\,\,  z_C =  h(\theta)\, \sin \theta + f^*(\theta)\, \cos \theta \,\,\,,\,\, f^*(\theta) = f_o - f(\theta) .
\end{equation}

We will use as  generalized  coordinates $(\theta , \psi ; \,x ,\, y, \phi)$ where $P = (x,y)$ is the contact point in the plane,  rather than using the  projection $T = (x_C, y_C)$ of the center of mass.  
The kinetic energy  $T =  T_{linear} + T_{ang} $   of the system is composed by 
\begin{equation}  \label{kinetic}
\begin{split}
& T_{linear} = \,\frac{1}{2} \, m\, \left[\dot{x}_C^2 + \dot{y}_C^2 + \dot{\theta}^2  \,  (dz_C/d \theta)^2  \right ], \,\,\, \, dz_C/d \theta = \Lambda(\theta) \\
& T_{ang} = \frac{I_1}{2} \left( \dot{\theta}^2 + \dot{\phi}^2 \sin^2 \theta \right) + \frac{I_3}{2} \left( \dot{\psi} + \dot{\phi} \cos \theta    \right)^2  \,\,\,
\end{split} \end{equation}
\begin{rem}
We borrowed the expression for the  rotational energy  $T_{ang}$ relative to the center of mass  $C$ from Arnold's book \cite{Arnoldbook}, section 30, pg. 151;  this  formula  is special for bodies of revolution.  It can be obtained directly  using (\ref{Omegaformula})  and the formula $T_{ang}= \frac{1}{2} (\Omega,I\Omega)$.
 Thus we avoid  changing the tensor or inertia from the center of mass $C$ to the contact point $P$. 
  \end{rem}
  
Thus to get the Lagrangian we  need only  to compute  $\dot{x}_C^2 + \dot{y}_C^2$.  But this is also in hands.

\begin{prp}  Since $(x_C, y_C) = (x,y) + \Lambda(\theta)\, e_N^\perp$   (see (\ref{Lambdadfn}))  then
\begin{equation} 
\begin{split}  \label{dotT}
& (\dot{x}_C, \dot{y}_C) = (\dot{x},\, \dot{y}) -  \Lambda(\theta) \, e_N \, \dot{\phi}  +   \frac{d \Lambda}{d\theta}\, e_N^\perp\, \dot{\theta} 
\end{split}
\end{equation}
\end{prp}

\begin{exan} \label{torusexa2}
Returning  to the torus,  
   $ h = R + r \,\sin \theta, \,\,  f_o -  f = r \,  \cos \theta,  \, r(\theta) = r . $ 
 Then  looking at Fig.5 \ref{fig:fig5} confirms the short calculation with Proposition \ref{PTprop}:  
\begin{equation}
(x_C, y_C) = (x, y) + R \cos \theta \, e_N^\perp
\end{equation}
 \end{exan}

Since the Lagrangian is $\phi$-invariant, we can do all the calculations with $\phi=0$:
\begin{equation} \label{dotxcyc} 
\begin{split}
&  \dot{x}_C  =  \dot{x} - \Lambda(\theta)\, \dot{\phi}\\
& \dot{y}_C  =  \dot{y}  +  \frac{d\Lambda}{d\theta} \, \dot{\theta}\
\end{split}
 \end{equation}
 We do not bother to collect  everything  now.  We will soon substitute the constraints (with $\phi=0$)  to obtain the compressed  Hamiltonian and the almost symplectic structure.  We will now make a digression to summarize the  reduction  theory.

 \section{Almost symplectic approach for Chaplygin reduction} \label{review}

Summary: The reduction  procedure for Chaplygin systems  in a  principal $G$-bundle $G \hookrightarrow Q \mapsto M$  yields   an almost Hamiltonian system in $T^*M$  with   almost symplectic form:     the canonical  2-form  plus 
 a semi-basic 2-form $J \cdot K$. Here    $J$ is the momentum map  of the group action,  and $K$ the curvature of the connection.
 In the case of rubber rolling $K$    is  well known by control theorists (Appendix \ref{controlK}). For a  surface  of revolution,    a conserved quantity arises, the reduced system is conformally Hamiltonian
   and can be  reduced to a 1 degree of freedom  Hamiltonian in   $(\theta, p_\theta)$.

\subsection{Reduction of  Chaplygin systems to the base manifold  }  \label{Review}

   A generalized Chaplygin system consists of a principal bundle $Q \rightarrow B$, with $M = Q/G$ being  the quotient of a group  $G$ acting on $Q$,  $G$-invariant Lagrangian $L(q,\dot{q}) $  and a  connection 1-form. Its      horizontal distribution ${\cal H} \subset 
 TQ$  gives the  constraints.   When $G$ is abelian one has  the classical Chaplygin systems (\cite{Chaplygin2002} 1897,  \cite{Chaplygin2008}).   We refer to  \cite{NeimarkFufaev} and \cite{Borisov2016a} for historical developments of nonholonomic systems.  For background on principal bundles and connections   we refer to  Warner's \cite{Warnerbook} and for  symplectic geometry  to Arnold's \cite{Arnoldbook}.  
 
\subsection*{The almost symplectic  procedure} \label{shortreview}

 We now summarize the theory presented in  \cite{Ehlersetal2005},  which is the Hamiltonian version of earlier results in \cite{Koiller1992} that were done  in the Lagrangian setting.  
Reduction   to   $ T^*M$ is done  
 as follows:

  The dynamics will be governed by  $(H_{red}, \Omega_{NH}) $  with   $H_{red}$   the `compressed'   Hamiltonian and  $\Omega_{NH} $ an almost symplectic form in $T^*B$
  \begin {equation} \label{JK}  \Omega_{NH}  =  \Omega_{can} +   J  \cdot  K .
  \end{equation}
  where $J$ is the momenum map of the action  and $K$ is the curvature of the connection. Step-by-step:

  \begin{enumerate}
  \item  \label{step1} %
 Do the  Legendre transform of $L$:    $V_q \in TQ \mapsto  P_q \in T^*Q$. 
  \item \label{step2} Construct the `compressed'  Lagrangian $L_{red}(b, \dot{b}) $ in $TM$ using any section of the bundle, and then  the  corresponding   Legendre transform
  $ TM \mapsto T^*M $ and  the reduced Hamiltonian $H_{red}$.   
\item  \label{step3} Form the pairing $J \cdot K$ which is  a semi-basic 2-form in $T^*M$, that we call ``pseudo-gyroscopic''.  
\end{enumerate} 
     
\paragraph{The clockwise diagram.}  Starting on  $p_m \in T^*M$, for any chosen $q$  on the fiber  of $Q$ over $m \in MB$,  we go clockwise to  $P_q \in Leg({\cal H}) \subset T^*Q$, to complete the missing arrow\footnote{A mathematical mantra: ``all diagrams should commute".}.

 \begin{equation} \begin{array}{lll}  \label{diagram} 
V_q  \in  {\cal H} \subset TQ  \,\,\,\,\,\,\,\,&\underset{Legendre}{ \longrightarrow} &   P_q \in Leg({\cal H}) \subset T^*Q \\
\, \uparrow & & \,\,  \\
\, hor & &\,\,  \\
\,\,  | &         & \,\,   \\
& & \\
v_m \in TM  & \underset{ {(Leg_{compr}})^{-1}}{\longleftarrow} \,\,\,\,\, &  p_m \in T^*M 
\end{array}
\end{equation}
\vspace{2mm}

 $J$ will be calculated at  $P_q$  producing an element of the  dual Lie algebra $Lie(G)^*$.  
The curvature $K$  is an $Ad_G$ invariant 2-form on $Q$ with $Lie(G)$
  values.  
The  pairing  $J \cdot K$ produces a semi-basic 2 form on $T^*M$ because    $J$ is $Ad^*_G$  invariant and the ambiguities cancel each other.  

This abstract nonsense  works quite  well algorithmically. 
If $r_i ,  p_{r_i} $   are conjugate coordinates in $T^*M$, 
the almost symplectic form writes as
\begin{equation} \label{almostNH}
\Omega_{NH} = \sum_{i=1}^m \, d p_{r_i}  \wedge dr_i +  \underbrace{\sum_{i<j}  \,  \sum_{k=1}^m  p_{r_k} \, f^{ij}_k(r) \, dr_i    \wedge dr_j}_{J \cdot K}    .
\end{equation}
 
Let us just elaborate  a bit  on the semi-basic two form $ J \cdot K$ for the more exigent reader. 
   Given two vectors $W_1, W_2 \in T_{p_m}( T^*M)$  one computes  (the real valued)    $ (J \cdot K) (W_1, W_2) $  as follows:
 \begin{itemize}
 \item  Project  $ W_i  \in T_{p_m}( T^*M ),\, i=1,2 $  to  $ w_i \in T_m M$ via the derivative of  
 $T^*M  \rightarrow M$.
 \item  Choose \textit{any}  element  $q$ in the fiber of  $m$ of the bundle $Q \rightarrow M$.
 \item Lift the vectors $w_i \in T_m  M$  to  horizontal vectors $hor(w_i) \in  T_qQ$.
 \item Compute  $K_q(h(w_1), h(w_2)) \in Lie(G)$.  Then ($J \cdot K) (W_1, W_2) =  J(P_q) (K_q(h(w_1), h(w_2))$.  
 \end{itemize}
 
  In some cases a  volume form on $M$ gives a   geometric meaning for  $J \cdot K$.    Appendix \ref{referee1} has some  recent informations, gently given by  one of the referees.
  \vspace{2mm}

\subsection{Momentum map $J$ of the $SE(2)$ action on   $Q =     SE(2) \times \Sigma \rightarrow \Sigma$}

 As I mentioned right from the beginning,  the  no-twist  
  together with the no-slip   constraints makes  rubber rolling a non-abelian Chaplygin system. 
 The bundle  is  a  trivial product, so  to compute the momentum $J$ I   just  took   the left  action  of $SE(2)$ on itself.
 
   By equivariance,  calculations  can (and will)  be done at the identity element  of the group (the simplest section of the bundle), which for $SE(2)$ is  $x=y=\phi=0$.  
 \vspace{1mm} 
 
[ Recall that the momentum map of a left action of a Lie group on itself is  $J: T^*G \rightarrow Lie^*(G)$ is
$ J(p_g) = R_{g^{-1}}^* p_g$, since
 $
 ( J(p_g)\,,\, \xi) = p_g ( \frac{d}{dt}_{|t=0} \, e^{t \xi}\, g)  =  p_g ( (R_g)_* \xi) =  ( (R_{g})^* p_g , \xi).  
 $ ].
 \vspace{2mm}
 
Since  at  the identity of the group the momentum map is the identity function\footnote{Thus we do not  use the  group multiplication 
  $(r,s,\alpha) (x,y,\phi) = (r + x \,\cos \alpha - y \, \sin \alpha , s + x  \sin \alpha  + y \, \cos \alpha, \phi + \alpha).$ }, then at the identity
 $  P = P_x dx + P_y dy + P_\phi d\phi  \in se(2)^* \,\,,\,\,\, \xi =\xi_x \, \frac{\partial}{\partial x} +  \xi_y \, \frac{\partial}{\partial y}  +  \xi_\phi  \, \frac{\partial}{\partial \phi }  \in se(2) \,\, $  one has simply
\begin{equation}
  (J(P) , \xi) = P_x \, \xi_x + P_y \, \xi_y + P_\phi  \, \xi_\phi  .
  \end{equation}

\begin{rem}  In the pairing $J \cdot K$, the factor $K$ is   pure geometry;  the factor $J$ carries the material  (physical) information via the Legendre transformation.
\end{rem}
 For a general convex  body (not necessarily of revolution),  
  $J \cdot K$  will be abbreviated as
\begin{equation} \label{abbrev}
\jk (\theta ,  \psi) = P_\phi(\theta, \psi) \, \sin \theta = 
( p_\theta \, f_1(\theta,\psi) +  p_\psi \, f_2(\theta,\psi))\, \sin \theta \,\,\,\, \,\,\,\, \text{(omitting}\,\,  d\theta \wedge d\psi) 
\end{equation}
The $\sin \theta$  comes from the curvature of the connection as explained by Proposition \ref{curvatureform1} to be  proven in next subsection.   The important  function 
\begin{equation} \label{impfun}
P_\phi(\theta,\psi) = p_\theta \, f_1(\theta,\psi) +  p_\psi \, f_2(\theta,\psi)
\end{equation}  
will be computed in section \ref{rubberroll}.   For surfaces of revolution   $f_1 \equiv 0$ and $f_2$ does not depend on $\psi$. 
\vspace{2mm}

\begin{prp}
 For a general convex  body (not necessarily of revolution) the equations of motion   with reduced Hamiltonian $H(p_\theta, p_\psi, \theta, \psi)$  will be of the form
\begin{equation} \label{reduce2dof}
\begin{split}
& \dot{\theta} =   H_{p_\theta}  \,\,\,,\,\,\,
 \dot{p}_\theta =   - H_\theta + \jk(\theta,\psi)\,  \dot{\psi} \\
& \dot{\psi} =   H_{p_\psi}    \,\,\,,\,\,\,
 \dot{p}_\psi  =       - H_\psi  -   \jk(\theta,\psi)  \, \dot{\theta}    \\
\end{split}
\end{equation}
The pseudo-gyroscopic terms do not affect  conservation of $H$.  
\end{prp}

\vspace{2mm}

\subsection{The curvature 2-form $K$ of the connection and the pairing  with $J$}

Recall the  geometric meaning  of the curvature of a principal bundle connection (see Warner's \cite{Warnerbook}): take the  lift  of a small  closed curve of the base $B$ to $Q$, with initial points $b_o \in B $ and $q_o \in Q$ respectively. 
  When one returns to $b_o$  in the base one reaches a point $q_1$ in the same fiber over $b_o$, which  in general is different from  $q_o$. 
  
If this is done infinitesimally, the jump in the fiber  is measured by an element of $Lie(G)$,   extracting the vertical component of   Lie brackets of horizontal lifts of two vectors in the base.

In the case of surfaces of revolution, the horizontal lifts to the total space  of  the tangent  vectors  in the Poisson sphere  $\partial_\theta=\partial/\partial_\theta $ and $\partial_\psi=\partial/ \partial_\psi$  are given by 
\begin{equation}   \label{lifts}
  h_\theta = \partial_\theta - r(\theta) \, e_N^\perp\,\,\,\,,\,\,\, h_\psi =  \partial_\psi - h(\theta) e_N \,   - \cos \theta \, \partial_\phi\, \,\, 
  \end{equation}
 
\begin{miracle} (This, an easy one).  
\begin{equation}
[h_\theta , h_\psi ] =  \sin \theta \, \partial_\phi  +  (\cancel{r(\theta) \cos \theta} \, -  \cancel{dh/d\theta})\,   e_N =    \sin \theta \, \partial_\phi 
\end{equation}
\end{miracle} 
This miraculous cancelation is due to  (\ref{miracle}).   
The Lie algebra elements   $\partial_x, \, \partial_y$  do not appear\footnote{Consider a closed curve in the torus bounding a simply connected domain. How about  the curved  in the plane?}.
  By   the definition of curvature  of a principal connection we have 
  \begin{prp}  \label{curvatureform1}
  For a general rubber rolling convex surface 
\begin{equation}  \label{curvatureform}
\begin{split}
& K = \, \,  \  \frac{\partial}{\partial \phi } \,  \sin \theta  \, d\psi   \wedge   d\theta    =  \,   \frac{\partial}{\partial \phi } \,\, dA_{S_{opposite}^2}  \\
& J \cdot K = -  P_\phi(\theta, \psi)  \,\,  \sin \theta \,\, d\theta \wedge d\psi .
\end{split}
\end{equation}
\end{prp}
\vspace{2mm}

\begin{rem}
The  result  in Appendix \ref{controlK} shows that it is not necessary to assume that the surface is of revolution.  What is subtle is the orientation for 
the area form. Since the Poisson vector is external, it is opposite to  $e_z$ at the contact.  This is why  we took   $\sin \theta  \, d\psi   \wedge   d\theta $ instead of $ \sin \theta  \, d\theta   \wedge   d\psi $. 
We insert  therefore a minus sign on the next proposition.  But we agree that that was a bit tricky.
\end{rem}
\noindent {\bf Confession.}  I perceived this only  \textit{a posteriori,} when  I was doing  the necessary adjustments  to obtain the same results of the Izhevsk group. \\
  
 \newpage

\section{Rubber rolling of bodies of revolution.  ``Nose'' function $N(\theta)$.  }  \label{rubberroll}

I will  now construct  the function  $  P_\phi(\theta, \psi) = p_\theta\,  f_1(\theta,\psi) +  p_\psi\,  f_2(\theta,\psi) $  in (\ref{abbrev}).   The coordinates  $p_\theta ,\, p_\psi, \theta, \psi$  are   canonical conjugate   in the    cotangent bundle of the sphere of Poisson vectors $\gamma \in S^2$.  For the reduced Hamiltonian, the Legendre transformation
from $ TS^2$ to $T^*S^2$  will be given in section \ref{compressedleg} . A body of revolution  has an  internal  symmetry and 
 $\psi$ will be ignorable:  $\, H_\psi \equiv 0.  $  Moreover
 the coefficient $f_1$  of $p_\theta$    in (\ref{abbrev})  vanishes,
\begin{equation} \label{claim1}
\jk = -  p_\psi \,  n(\theta)   \,\,\,, \,\, \,  n(\theta)  =   f_2(\theta)\, \sin \theta 
\end{equation}
  As it is to be expected,  this leads  to  a conserved quantity $\ell$ and the  system (\ref{reduce2dof})  will  be  further  reduced   to  a 1 degree of freedom Hamiltonian in $(\theta, p_\theta)$.  The following  expression is key:  

\begin{dfn} The ``Nose" function:
\begin{equation} N(\theta) =  \exp( \int \, n(\theta) \, d\theta )
\end{equation}
\end{dfn}
\smallskip

  At this point I became very  frightened.  In concrete examples,  would the analytical calculation  of this integral become intractable?  To my  amazement,   in rubber rolling  of a surface of revolution the calculation  was  a `free lunch'.   Why is so?   A mystery, but a fortunate one.  
  \begin{rem} About strange occurrences,  in Gogol's short story,   the   Nose   says to Collegiate Assessor Kovalyov, when they met for the first time (after certain  day he found himself  to be Noseless):
\begin{quote}
``I exist in my own right. Besides, there
can be no close relation between us. Judging by the buttons on your
uniform, you must be employed in the Senate or at least in the Ministry
of Justice. As for me, I am in the scholarly line."  
\end{quote}
\end{rem}

\begin{prp} \label{gandG} $\,$   

  \begin{enumerate}[i)]  
\item  
Conserved quantity
\begin{equation}  \ell =  \frac{p_\psi}{N(\theta)} \,\,\,, \,\,\, N(\theta) = \exp(\, \int n(\theta) \, d\theta ) .
\end{equation}
\item The almost symplectic 2-form  (\ref{almostNH}) in the base $T^*S^2$ 
 \begin{equation}
\Omega_{NH} = dp_\theta \wedge d\theta + dp_\psi \wedge d\psi  - p_\psi \,n(\theta) \, d\theta \wedge d\psi
 \end{equation}
is conformally symplectic:
  $ \text{let}\,\, \tilde{\Omega} =  N^{-1}  \, \Omega_{NH}.  \,\,\, \,\, \text{Then} \,\,\,   d \tilde{\Omega} = 0 .$ 
\end{enumerate}
\end{prp}

\noindent  {\bf Proof.}  For i) just  integrate the  last equation in (\ref{reduce2dof}): 
$ \,\,\, \dot{p}_\psi  -   p_\psi \, n(\theta) \dot{\theta} = 0 $.  \,\,\,   ii)  is standard for experts, the novice reader should not have any trouble also.

For any function $H$ on the base, let  the vector field  $X_H$ be  defined by $\Omega_{NH}(X_H, \bullet) = dH(\bullet)$. Then the vectorfied $N  X_H$ is hamiltonian for $\tilde{\Omega}$. 
Under a  \textit{temporary} time change of coordinates,
\begin{equation} \label{change1}
dt  =  N(\theta)  \, d\tau
\end{equation}
the reduced Chaplygin system in $T^* S^2$ is Hamiltonized:   
one can safely  do the  Marsden-Weinstein reduction. 
The infinitesimal generator of the $S^1$ action $\psi \mapsto \psi + \epsilon$ is obviously  $X = \partial/\partial \psi$, and  by definition of  momentum map $J$,
$$
 \tilde{\Omega}(X, \bullet) = dJ(\bullet).
$$ 
The obvious \textit{ansatz}  for the momentum map  is   $ J =  \ell =  p_\psi/N(\theta)  $.

\subsection*{Marsden-Weinstein reduction to 1 DoF} \label{MW}

According to the MW recipe, the reduced symplectic manifold is
$$
M = \frac{\{ (p_\theta, \theta , p_\psi , \psi  \, | \,  p_\psi = \ell\,  N(\theta)   \}} {  \{ \text{the ignorable}\,   \psi  \,  \} }
$$
It  is amusing to compute the reduced symplectic form as a ``sanity check":
$$
\Omega_{red} = N^{-1}  \, \left( dp_\theta \wedge d\theta + \cancel{d(\ell N) \wedge d\psi } -  \cancel { (\ell  \, N)  \, n \,\, d\theta \wedge d\psi }    \right) \,\,\,\, (\text{recall}\,\,\, n = N'/N)
$$
\textit{Now one   neglects the conformal factor $N$ in front of $\Omega_{red}$,  thus   returning  to the original time   $t$. }

\begin{prp} \label{prop1DoF}
The reduced system to 1 DoF   stays  in the original time scale and   has the usual symplectic form  $dp_\theta \wedge d\theta$ with 
\begin{equation} \dot{\theta} =    \frac{\partial \tilde{H}}{\partial p_\theta }    \,\,\,\,,     \,\,\,\,\,\, \dot{p}_\theta =  -  \frac{\partial \tilde{H}}{\partial \theta } \,\,\,,\,\,\, \tilde{H} = H(p_\theta, \theta ;  p_\psi)_{|\,  p_\psi = \ell N(\theta) } 
\end{equation}
 where $\ell$ is the  constant of integration,  $ \ell = p_\psi/N(\theta)$.
\end{prp}

\noindent {\bf  Note.}  The  reviewer asked  me to clarify why the time change     (\ref{change1})  was  done just to be undone. 
Perhaps   the  time change was an excess of  care.  In hindsight the reduction  to 1 DoF can be done in the NH framework  in  the same way as the Marsden-Weinstein reduction in the     
  Hamiltonian setting.
  With the danger of being redundant, the formal steps are:

\begin{enumerate}[i)]
\item The 2-form  $ \tilde{\Omega} = N^{-1} \, \Omega_{NH} $ is symplectic.
\item Consider a  vectorfield $X_H$ satisfying  $ \Omega_{NH} (X_H, \bullet) = dH(\bullet)$ for a given function $H$ on the cotangent space of the base (here $S^2$).  Clearly
$$ \tilde{\Omega}( N X_H , \bullet) =  \frac{1}{\cancel{N} } \, \Omega(\cancel{N} \, X_H , \bullet) = dH . 
$$
\item This means that $N X_H$ is\textit{ bona fide Hamiltonian  vectorfield } for $\tilde{\Omega}$ with Hamiltonian $H$, and in our case $\psi$ is ignorable.
\item In other words, the MW reduction will give
$$
\cancel{N^{-1}}  \, \left( dp_\theta \wedge  d\theta \right) [ ( \cancel{N} X_H)_{\text{reduced to 1 DoF}} , \bullet ] = d \tilde{H}\,,\,\,\, \tilde{H} = H(p_\theta, \theta ;  p_\psi)_{|\,  p_\psi = \ell N(\theta) } 
$$

\end{enumerate}

\vspace{2mm}

 We are ready for  the calculations.  
 Miracles will happen.

\subsection {Calculation of  $P_\phi = \partial  L/\partial \dot{\phi}$ at $\phi = 0$}

I need this calculation  for the $J \cdot K$ term, see  (\ref{abbrev}).  
One   looks  where  $\dot{\phi}$ appears  in   the expressions  of the skidding Lagrangian in section \ref{skidding1}  (always with $\phi = 0$). 
I got  readily 
$$
P_\phi =  I_1 \, \sin^2 \theta \, \dot{\phi} + I_3 \, (\dot{\psi} + \dot{\phi} \, \cos \theta) \, \cos \theta -
m\, \Lambda(\theta)\,  (\dot{x} - \Lambda(\theta)\,  \dot{\phi}) .
$$
Then the no-twist   
 and the  no-slip  condition  for $\phi = 0$ are inserted: $\dot{\phi} = - \cos \theta \, \dot{\psi} \,\,,\,\,\,  \dot{x} = - h(\theta) \, \dot{\psi}$.   
\begin{equation*}
P_\phi  = \left\{ (I_3-I_1)\, \sin^2 \theta \, \cos \theta  +  m\, \Lambda(\theta) \, [  \, h(\theta) - \Lambda(\theta) \, \cos \theta] \, \right\}\, \dot{\psi} 
\end{equation*} 
Now I  recall from Proposition \ref{PTprop} that 
$ h - \Lambda \, \cos \theta  = z_C(\theta)  \, \sin \theta . $
\begin{prp}
\begin{equation}
P_\phi = C(\theta) \, \dot{\psi}\,\,,\,\, C =  \left\{ (I_3-I_1)\, \sin  \theta  \cos \theta +  m\, \Lambda(\theta) \, z_C(\theta)   \, \right\}\, \sin \theta
\end{equation}
\end{prp}

\begin{exan}
For the torus (see Example \ref{torusexa}), we know  $h = R + r\sin\theta,\, \Lambda =  R \cos \theta,\, z_C = r + R\, \sin \theta$  (see (\ref{Lambdatorus}) hence 
\begin{equation}
 P_\phi = C(\theta)\, \dot{\psi} \,\,\,,\,\,\,\, C =  \left[ (I_3 - I_1 + m \, R^2 )\, \sin \theta + m \, R\, r \,\right] \, \sin \theta \cos \theta  
\end{equation} 
\end{exan} 

\subsection{The compressed Lagrangian  with $\phi = 0$}  \label{compressedleg} 

I will now insert the constraints in the   `skidding'  Lagrangian.   
  The contribution from the angular kinetic energy  (\ref{kinetic})  involves the inertia coefficients $I_1,\, I_2$
  $$
  T_{ang} = \frac{1}{2}\, I_1 \, \dot{\theta}^2 +  \frac{1}{2}\, \sin^2 \theta \, ( I_1 \cos^2 \theta + I_3 \sin^2 \theta)\, \dot{\psi}^2  .
  $$
  For  the linear kinetic energy, there will be right away the contribution  from the $\dot{z}_C^2 = \Lambda^2(\theta) \, \dot{\theta}^2 $. 
  Finally,  one  must add the contribution from  $\frac{1}{2}\, m\, (\dot{x}_C^2 + \dot{y}_C^2)$.  From (\ref{noslip}) and (\ref{dotT}) I got 
  $$
 \dot{x}_C = \dot{x} - \Lambda \, \dot{\phi}\,\,\,\,\text{with} \,\, \,\, \dot{x} = - h\, \dot{\psi} \,\,\,; \,\,\,\, 
 \dot{y}_C = \dot{y} +  \frac{d\Lambda}{d\theta}  \, \dot{\theta}\,\,\,\,\text{with} \,\, \,\,  \dot{y} = - r(\theta)\, \dot{\theta}.
  $$
 Hence
 \begin{equation}
  \frac{1}{2}\, m\, (\dot{x}_C^2 + \dot{y}_C^2) =   \frac{1}{2}\, m\, \left[(r(\theta) -  \frac{d\Lambda}{d\theta})^2\,  \dot{\theta}^2 + (\underbrace{h - \Lambda \, \cos \theta}_{z_C \sin \theta})^2 \, \dot{\psi}^2             \right]
  \end{equation}
\begin{prp} \label{ABprop}  We have  $T_{compr}  = \frac{1}{2} ( A(\theta)\,  \dot{\psi}^2 + B(\theta) \, \dot{\theta}^2 )$ 
with 
\begin{equation} \label{AB}
\begin{split}
& \,\,\,\,\,\,\, A(\theta) = \underbrace{( I_1\, \cos^2 \theta + I_3\, \sin^2 \theta + m \, z_C^2(\theta) )}_{N^2(\theta)}  \, \sin^2  \theta    \\
& \,\,\,\, \,\,\, B(\theta) =  I_1 + m \, \left[ \underbrace{ \Lambda^2    +  \, (r(\theta) - d\Lambda/d\theta)^2}_{\text{see  Proposition \ref{PTprop}} } \right] = I_1 + m \,  |CP|^2
 \end{split}
\end{equation}
\end{prp}
\begin{rem}
Note that $B(\theta)$ does not depend on $I_3$.   The fact that  Proposition \ref{PTprop}, equation (\ref{dLambda}) can be used  yielding  $|CP|^2$ was a pleasant miracle.
\end{rem}

\begin{exan} For the torus,  $ \Lambda = R \, \cos \theta\,,\, h = R + r \, \sin \theta \,\,,\,\,  r(\theta) = r $ hence
\begin{equation}
\begin{split}
& A =  \left[  I_1\, \cos^2 \theta + I_3 \, \sin^2 \theta + m ( r + R\, \sin \theta)^2   \right] \, \sin^2 \theta \\
& B = I_1 + m ( R^2 + r^2 + 2 r\, R \, \sin \theta )
\end{split}
\end{equation}
\end{exan}

\subsection{Computing the conserved quantity.  The big  miracle.}  

We start with the torus example.  Coming back to Proposition \ref{gandG}  
 we can now write down (\ref{claim1}),
\begin{equation}
n =  \frac{C}{A} \sin \theta 
 =   \frac{  \left[ (I_3 - I_1 + m \, R^2 )\, \sin \theta + m \, R\, r \,\right] \, \cancel{\sin \theta}  \cos \theta  
 }{\left[  I_1\, \cos^2 \theta + I_3 \, \sin^2 \theta + m ( r + R\, \sin \theta)^2   \right] \, \cancel{\sin^2 \theta }}\, \cancel{ \sin \theta} 
\end{equation}

We must compute the indefinite integral
$$
\int n(\theta) \, d\theta =  \int \,  \frac{  \left[ (I_3 - I_1 + m \, R^2 )\, \sin \theta + m \, R\, r \,\right] \,   \cos \theta  
 }{\left[  I_1\, \cos^2 \theta + I_3 \, \sin^2 \theta + m ( r + R\, \sin \theta)^2   \right] \,}\,   \, d\theta
$$
or making $x = \sin \theta$:
$$
\int \, n(\theta)\, d\theta    = \int \,  \frac{  \left[ (I_3 - I_1 + m \, R^2 )\, x + m \, R\, r \,\right] \, 
 }{\left[ (I_3 -I_1) \, x^2  + m ( r + R\, x)^2  + I_1  \right] \,}\,    dx
$$

\begin{miracle} We observe that very nicely, the numerator is 1/2 of the derivative of the denominator.
\end{miracle} 
\textit{ Thus without any calculation, }
\begin{equation} \label{G1}
N(x) = \exp(\int n) =  \sqrt{ (I_3 -I_1) \, x^2  + m ( r + R\, x)^2  + I_1 }
\end{equation}
This  very nice  phenomenon is general:  since $\Lambda = dz_C/d\theta$  (Proposition \ref{dzC}) we can write
\begin{equation} \label{truemiracle}
\int \, n \, d\theta = \int\, \frac{C}{A} \sin \theta \, d\theta =  \int\, \frac{\left\{ (I_3-I_1)\, \sin  \theta  \cos \theta +  m\, \Lambda(\theta) \, z_C(\theta)   \, \right\}\, \cancel{ \sin \theta}
 }{( I_1\, \cos^2 \theta + I_3\, \sin^2 \theta + m \, z_C^2 ) \, \cancel{\sin^2  \theta}   } \, \cancel{\sin \theta}   d\theta
\end{equation}
and indeed \textit{the derivative of the denominator is twice the numerator. }
\vspace{2mm}

\begin{prp} For any  surface of revolution rubber rolling over the plane, after the  the Chaplygin reduction for $(\theta, \psi , p_\theta , p_\psi)$  we have the conserved quantity  $\ell =  p_\psi/N(\theta)\,  $, with 
\begin{equation} \label{G2} 
N  =   \sqrt{ I_1\, \cos^2 \theta + I_3\, \sin^2 \theta + m \, z_C^2(\theta) } .
\end{equation}
\end{prp}
\begin{rem}
Note that $A(\theta) = N^2(\theta) \sin^2 \theta$  so 
\begin{equation} \label{ufaufa}
\ell =  (N^2 \, \sin^2 \theta\,  \dot{\psi}) /N(\theta) =  N(\theta)  \, \underbrace{\sin^2 \theta\, \,\, \dot{\psi}}_{\Omega_3\, (\text{see  (\ref{ufa})})} =  N(\theta) \, \Omega_3
\end{equation}
 which coincides with Eq. (5.1) in \cite{Borisov2008}.
\end{rem}

\begin{miraclen} 
Since  $$H_{red} = \frac{1}{2} \left( \frac{p^2_\theta}{B(\theta)} +  \frac{p^2_\psi}{A(\theta)}     \right)$$   and $p_\psi = \ell \, N(\theta)$  we have
$$
\frac{p^2_\psi}{A(\theta)} =   \frac{\ell^2 \, \cancel{N^2(\theta)}}{ \cancel{N^2(\theta)}  \, \sin^2 \theta}  = 
 \frac{\ell^2}{\sin^2 \theta} 
$$
\smallskip

The dependence on $A(\theta)$ disappears, and with it the dependence on $I_3$.   
\end{miraclen}

\subsection{Main result:  reduced system  in $(\theta, p_\theta)$}

From Proposition \ref{prop1DoF}   
 in section  \ref{MW} one  can state:
\begin{teo} The   reduced Hamiltonian in  $(p_\theta, \theta)$ is
\begin{equation}
\tilde{H} = \frac{1}{2} \left( \frac{p^2_\theta}{B(\theta)} + \frac{\ell^2}{\sin^2 \theta}      \right) + m g\, z_C(\theta)\,\,\,,\,\, B =  I_1 + m \, |CP|^2 . 
\end{equation}
where $|CP|$ is the distance of the center of mass to the contact point  when the body axis is making an angle $\theta$ with the vertical. For the reconstruction of  $\psi(t)$  one uses  $\dot{\psi} =  \ell / [N(\theta(t)) \sin^2 \theta(t)] $  with 
$$N  =  \sqrt{ I_1\, \cos^2 \theta + I_3\, \sin^2 \theta + m \, z_C^2(\theta) },$$   and one uses the  the no-twist constraint to recover $\phi(t)$.   Finally the no-slip constraints allow to recover the planar motion $(x(t), y(t))$. 
\end{teo}

\begin{teo}(Kindly pointed out to us by Alexander Kilin).

The stretching in the $p_\theta$ direction defined by
\begin{equation}
\tilde{p}_\theta = \frac{p_\theta}{\sqrt{B(\theta)}}
\end{equation}
does not change qualitatively the  features of the level curves of the Hamiltonian, that becomes simply
\begin{equation} \label{tildeV} 
\tilde{H} = \frac{1}{2}\, \tilde{p}^2_\theta + \tilde{V}(\theta),\,\,\,\,\,\, \tilde{V} =  \frac{\ell^2}{2 \sin^2 \theta}     + m g\, z_C(\theta) 
\end{equation}
The reduced symplectic form $dp_\theta \wedge d\theta $  becomes
$$
dp_\theta \wedge d\theta =   \sqrt{B(\theta)}  \, d\tilde{p}_\theta \wedge d\theta .
$$
\end{teo}
This  means that under this other  time change of coordinates   
\begin{equation} \label{change2} 
dt = \sqrt{B} \, d\tau,
\end{equation}  
the final result is the 
classical 1 DoF motion under the potential $\tilde{V}$  in (\ref{tildeV}),  
\begin{equation}
\theta'' = - \frac{d \tilde{V}}{d \theta }\,\,\,   (  '= d/d\tau).
\end{equation}

If action-angle variables  $I, \alpha$  are constructed  for  a regime of solutions,  $\theta = \theta(I, \alpha),\,\, \tilde{p}_\theta = \tilde{p}_\theta(I, \alpha)$,  with  frequency 
$\omega(I)  = \partial \tilde{H}/\partial I$,  for $\tilde{H} = \tilde{H}(I),$ then the solution for $\theta(t) $  in this regime will be
\begin{equation}
\theta(\tau) =   \theta_o +  \omega(I)\, \tau \,\,\,,\,\,\,  t  = \int \, \frac{d\tau}{\sqrt{B(\theta(\tau))}} . 
\end{equation}

\begin{rem}
In \cite{Borisov2013} the functions  that we denote  $N$ and $B$ appear  in Eqs. (3.2)  as  $g_1 \equiv N^2, \, g_2 = B .$ 
\end{rem}

\noindent {\bf Equation for relative equilibria.}  The equilibria of the 1 DoF system  have  $p_\theta = 0$ and
\begin{equation} \frac{d z_C}{d\theta} =  \frac{\ell^2}{m g} \, \frac{\cos \theta}{\sin^3 \theta}
\end{equation}

\noindent {\bf Dimensionally.}
\begin{equation}
\begin{split}
&[A] = [B] = M L^2 \\
[&G] = \sqrt{M} \, L \,\,\, \,\,\, (\text{square root in dimension analysis by abuse)}\\
&[\ell = G \, \dot{\psi} ]  =  \sqrt{M} \, L/T  \,\,\,\,,\,\,\,\, 
 [\ell^2] = M L^2/T^2 = [m g R ] = \text{energy}   \end{split}
 \end{equation}

  \subsection{The torus} 

  For the torus, recall  $z_c = r + R \sin \theta$  and $|CP|^2 = R^2 + r^2 + 2 R r \sin \theta$. 
I found in  the web   the principal moments of  inertia of the torus  relative to the center of mass.     In both cases, $I_3 >  I_1$.  
 \begin{itemize}
 \item  For solid torus: $I_3 = \frac{m}{4}(4R^2 + 3r^2),\,\,\, I_1=I_2  =  \frac{m}{8}(4R^2 + 5r^2) $
\item  For hollow torus:  $I_3 = \frac{m}{2}(2R^2 + 3r^2),\,\,\, I_1=I_2  =  \frac{m}{4}(2R^2 + 5r^2) $
   \end{itemize}
The potential is 
\begin{equation}
\tilde{V} =  \frac{\ell^2}{2 \sin^2 \theta} + mg R \, \sin \theta +  m g\,  r 
\end{equation}

Follows just  some elementary observations. 
The problem has discrete symmetry about $\theta=\pi/2$.  The equation for the equilibria is
\begin{equation}
 \cos \theta =   \frac{\ell^2}{m g R}  \, \frac{\cos \theta}{\sin^3 \theta}\, .
\end{equation}

 \begin{figure}[h] \label{figV}
\centering
     \includegraphics[width=0.8\textwidth]{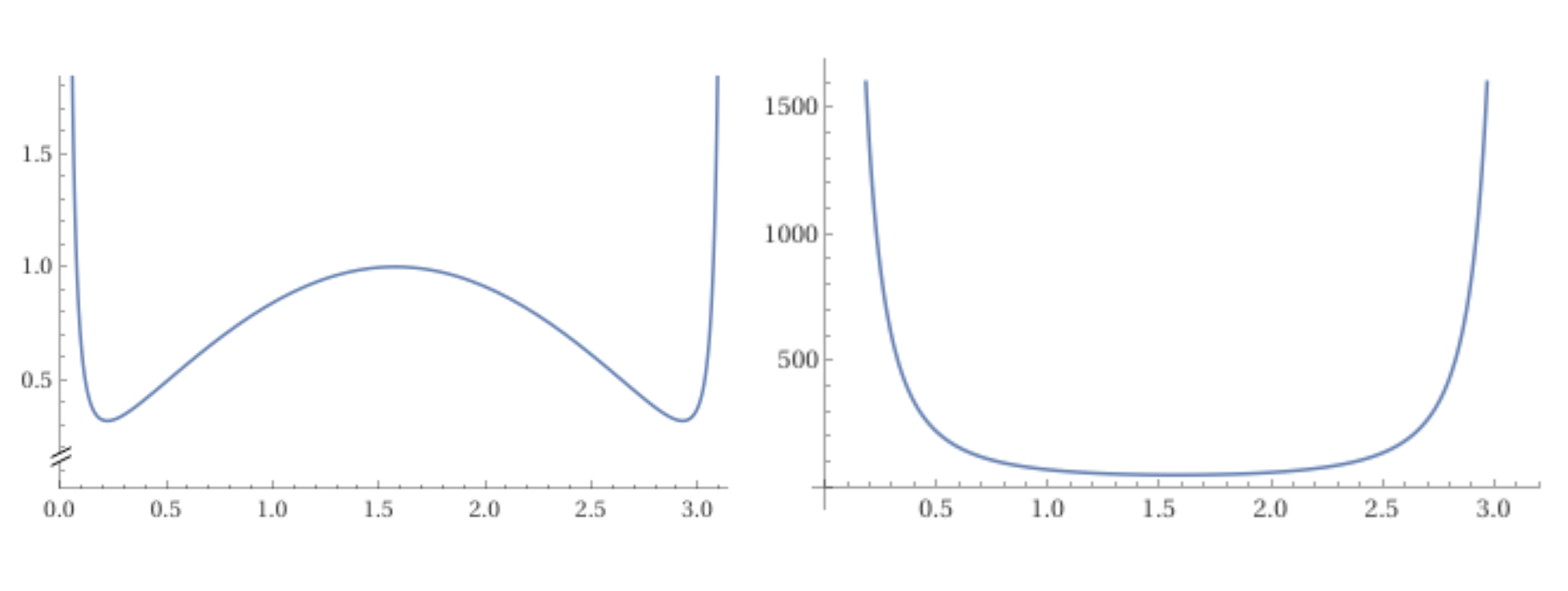}
\vspace{-8mm}
 \caption{\small{Plot of  $V = \ell^2/2\sin^2 \theta + \sin \theta,\,\,  (m=g=R=1,  \, \text{I  omit the constant})$   Left: $\ell = 0.1$.   When  $\ell^2 <m g R$  there are two relative stable equilibria satisfying  $\sin \theta =  (\frac{\ell^2}{m g R})^{1/3}$,  for which  the torus rolls over a circle in
 the plane. As 
 $\ell^2 \rightarrow  m g R$  the radius of this circle tends to infinity. The vertical position $(\theta=\pi/2)$  gives a separatrix between three regimes: the invariant manifolds correspond to the torus falling 
 along the line $e_N^\perp$ (in time reversal) , or roll along $e_N$ in direction  corresponding to one of the wells; Right: $\ell = 10$.  When  $\ell^2 > m g R$,   the vertical position of the torus  ( $e_3$  horizontal) is   stable.  In the body frame  turns around its axis
 very quickly ($\Omega_3 >> 1$).  For a very small amount of energy added to the minimum, $\theta$ remains close to $\pi/2$ and $\dot{\phi} \sim 0$. 
}}  
 \end{figure}

\subsection*{Immediate consequences  for the torus trajectories}

If    $\theta \neq  \pi/2$ the relative equilibria are
\begin{equation}
 \sin \theta =  (\ell^2/mgR)^{1/3} \,\,\,  \text{when}\,\,\, \ell^2 < mgR .
\end{equation}

\noindent {\bf Fig.6, left  ($\ell^2 <  mg R$).}  We have bi-stability in this case.  
 The  $\theta$-motion can be confined in one of the wells  or,   when the  energy is above that  of   
$\theta = \pi/2, $   to flip wildly between near $\theta = 0$ and  $\pi$.  
For  $\ell \sim 0$, although the equilibrium is near   $\theta_o \sim 0 $ or $\pi$,    a small amount of  energy  allows the torus  to lift  considerably.  This is reminiscent of what Cushman and Duistermaat found for the (marble) coin.  At any rate, when the  torus is almost flat,  it slowly rolls  over a circle in the plane with radius slightly bigger than $R$ and scalar velocity 
$ \sim R \ell/\sqrt{I_1 + m r^2} .$\\
 
\noindent {\bf Fig.6, right  ($\ell^2 >  mg R$).}     $\, \theta = \pi/2$  is a stable equilibrium for the 1 DoF system.  As the picture shows, it is quite flat. Adding some finite energy, the wobbling
in $\theta$ is wide.  
After a car crash,  a loose rubber tire can go far away.

\section{Some research directions and final comments  } \label{historical}  
 
\subsection{Analytic solutions}  \label{further}  

  \noindent    {\bf Problem.} 
  For the torus,   (\ref{tildeV}) lives on a  genus 2 Riemann surface.  Solving  with special functions, including  reconstruction is in order.   As an initial study, one could take the harmonic approximations to  $\theta(t)$ near the bottom of a well, so that the whole reconstruction could be done  (approximately)  with elementary functions. 
 A complete  analysis  for the rubber torus should be done in the same vein as  in \cite{KilinPivovarova2024}
    for the ellipsoid\footnote{Also with  movies/iterative  widgets.   Example: 
    \url{https://mathr.co.uk/blog/2013-04-18_rolling_torus.html}.}.   
 Specially challenging are the nearly flat motions.  
  
  \vspace{2mm} 
\noindent {\bf Historical note.} Analytic solutions for rolling coins were found by the `ancients'. 
  See  \cite{Borisov2003},    \cite{Batista2006}, \cite{Batista2008},    
 \cite{Cushmanetalbook2009}  specially chapter 7, for modern accounts.    
 As far as we know, the  nearly horizontal motions of rolling coins were studied  originally in  \cite{Cushman1996} and  \cite{CushmanDuistermaat2006}.  
  We quote  the abstract of the latter:
  \begin{quote} \vspace{- 3mm}
   {\footnotesize
``We study the motion of a disk which rolls on a horizontal plane under the influence of gravity, without slipping or loss
of energy due to friction. There is a codimension one semi-analytic subset $F$  of the phase space such that the disk falls
flat in a finite time, if and only if its initial phase point belongs to $F$. We describe the motion of the disk when it starts
at a point $p \ni  F$  which is close to a point $ f \in  F$.  It then almost falls flat, after which it rises up again. We prove that
during the short time interval that the disk is almost flat, the point of contact races around the rim of the disk from
a well-defined position at the end of falling to a well-defined position at the beginning of rising, where the increase of
the angle only depends on the mass distribution of the disk and the radius of the rim. The sign of the increase of the
angle depends on the side of $F$ from which $p$ approaches $f$"
}
   \end{quote}
  \newpage

\subsection{Adding dissipation}   \label{dissipation}

It is never   easy to construct models to match experiments.   
   I  think it is worth to present some  comments about friction  as Alexey Borisov and colleagues made a  bold step to start a laboratory\footnote{In the times of the industrial revolution the  focus was finding 
 laws for the  resistance to tangential motion between solids and liquids or gases. We quote from \cite{Lander1916}: ``A general relation between the dimensions of the surface, the velocity, the density of the fluid, and its viscosity bad been surmised as a consequence of the laws of motion by Stokes, Helmholtz, and Osborne Reynolds, but it was left to Lord Rayleigh to show, from the principle of dynamical similarity, that the phenomena involved could be expressed definitely by a simple mathematical formula".}. 
The practical importance of  studying rubber tires friction  is obvious from industrial and safety reasons.  I  did a quick search and found several experimental  works, e.g.  \cite{Pacejka1992}, \cite{Guo2006}, \cite{Heinrich2023}, \cite{Ejsmont2024}, \cite{Lenzi2024}.

    So far,  in  studying rolling friction,   mathematicians and physicists have  for the most part done numerical simulations not  yet  directly  connected with experiments, see eg. \cite{Bazzi2013} and \cite{Morinaga2014} for   tires, and \cite{Shamin2019} for the sleigh.
  Rayleigh dissipation principle is revisited in \cite{Yao2011}, \cite{Perez2018}, \cite{Gaset2020}. 
  Manuel de Leon and  collaborators look at external forces from the Geometric Mechanics viewpoint  \cite{Asiermaster}, \cite{Manolo2021}, \cite{Manolo2021a}, \cite{Manolo2022}, \cite{Manolo2022aa}, \cite{Colombo2022}. 
 Kilin and Pivovarova \cite{Kilin2021} investigated the interesting problem of    integrals of motion that  survive under dissipation, specially the Jellet-Chaplygin integral,  discussed in Borisov and Ivanov \cite{Borisov2020}, and even earlier in \cite{Ivanov2009}. A specific model  for dissipation in rolling systems was proposed by Ivanov \cite{Ivanov2019}.

Still controversial are the details of  final motions of rolling coins when dissipation is present\footnote{There is a commercial toy \url{https://toolsandtoys.net/eulers-disk-spinning-desk-toy/}, \url{http://www.eulersdisk.com/}.}.     These studies have been done only  for the marble constraints. 
 Some references for the  ringing  phenomenon are   \cite{Moffatt2000},  \cite{Kessler2002},  \cite{Petrie2002},  \cite{Caps2004}, \cite{Leine2008},  \cite{Abbott2010}, \cite{Baranyai2017}. See also  Borisov,  Mamaev and Karavaev \cite{Borisov2015}.  Anoher interesting phenomenon is the ``spiral reversal"  \cite{Jalali2016},   \cite{Borisov2017}.  Do similar  phenomena hold when no-twist is added?

  \subsection{Quick take on control and chaos} If the torus is slightly unbalanced or deformed, one expects   chaotic phenomena. See  \cite{Kazakov2013} for the  balanced, dynamically asymmetric ball. For the torus, a Melnikov splitting could be done at  the separatrix curve  of  the two wells. 
   Another important  research topic is  about \textit{control}, which is important for robotic applications.   For the rubber  sphere with $I_1=I_2 \neq I_3$ see \cite{Cendra2010}, and for different inertias
   \cite{Mamaev2020}. 
      The  literature about control of nonholonomic systems goes back at least 1997 \cite{Koon1997} and its interest is increasing, wee eg.   \cite{Manolo2010}, \cite{Bloch}, \cite{Kilinetal2015}.
      
 \smallskip
\newpage

\subsection{Can the  miracle in the ``Nose" function be explained? }

The miracle did not escape to Borisov and Mamaev in \cite{Borisov2008}.  They described  in the following way: 
\begin{quote}
``The fact that this integral  [$\ell = N(\theta) \Omega_3$] can be expressed in elementary functions for any body’s [of revolution] shape is far non-trivial. For example, in the problem of motion of a body of revolution on a plane when spinning is allowed this integral can be expressed in elementary functions only when the surface of the body is a sphere."     
\end{quote}

\noindent The pragmatic reader might accept  the Nose function   the same way as  Gogol  ends his story: 
\vspace{-1.5mm}  
\begin{quote}
``The most incomprehensible thing of all is, how authors can choose such subjects for their stories. That really surpasses my understanding. In the first place, no advantage results from it for the country; and in the second place, no harm results either.

All the same, when one reflects well, there really is something in the matter. Whatever may be said to the contrary, such cases do occur—rarely, it is true, but now and then actually." 
(\url{https://www.gutenberg.org/files/36238/36238.txt}) 
\end{quote}

\section{Final comments}

 I  have the  feeling that   Chaplygin's  reducing multiplier  method \cite{Chaplygin2008}  is related to $J \cdot K$ term 
 and  Hamiltonization, but we leave this just as loose question, up for grabs.

I worked here  with  very basic differential geometric  tools. I hope the reader may find it useful  or at least amusing.    I  advocate for doing  the reduction to $T^*M$  of  generalized Chaplygin systems in a principal bundle  $G \hookrightarrow Q \rightarrow M$   via 
the almost symplectic  approach, which can be done in  algorithmic fashion.

In principle the  almost symplectic implementation  could be done without drawing a single picture. As Lagrange boasted\footnote{However, Arnold's  \textit{Mathematical Methods of Classical Mechanics}, published 190 years after, has 269 illustrations in the translation to English by A.  Weinstein  and K. Vogtmann.   That is why  I did   some pictures   without feeling embarrassed.}:  \begin{quote} ``On ne trouvera point de Figures dans cet Ouvrage. Les méthodes que j'y expose ne demandent ni constructions, ni raisonnemens géométriques ou méchaniques, mais seulement des opérations algébriques, sujetties à une marche réguliere $\&$ uniform." (\textit{Avertissement} in  the beginning of his M\'ecanique Analitique (1788) \\

\end{quote}

\newpage

\noindent {\bf {Acknowledgements.}} 
  I  thank Alexander Kilin  and Elena Pivovarova  for explaining  the main features of the problem and  giving  references.  Alexander was very  patient to pointing  to us many distractions and typos in the various early versions I  sent to him.
   I  thank
 Alejandro Cabrera and Paula Balseiro  for several discussions, and to Luis Garc\'ia-Naranjo for allowing  me to use some of his comments as an Appendix.   \\

\noindent  {\bf I would like to add  a disclaimer. }   As I am no young anymore, I could use   a Chinese proverb   that  Prof. S. S. Chern said 
 (joking about himself) in a basic  Differential Geometry class that I  had the privilege to attend many years ago and  at that time I  could  not appreciate its wisdom:  \vspace{-2mm}
\begin{figure}[h]
\centering
   \includegraphics[width=0.25\textwidth]{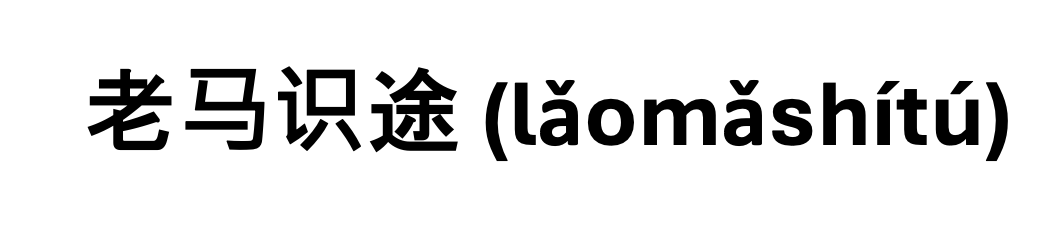}  \label{fig:chinese}
 \vspace{-7mm}  
\end{figure}

\centerline{\small{``Old horses know their way"} }

\vspace{2mm} 
\noindent  
However, in my case, in present time perhaps a Russian proverb would fit better: \vspace{-2mm}
\begin{figure}[h]
\centering
   \includegraphics[width=0.4\textwidth]{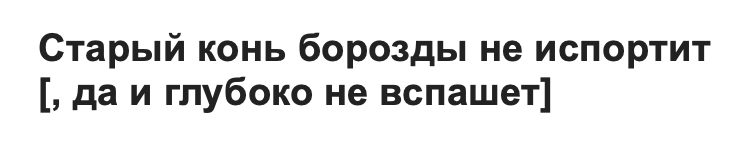}  \label{fig:russian}
 \vspace{-6mm} 
\end{figure}

\centerline{\small{``An old horse will not spoil the furrow,  but won't plough  a new one either".} }

 \vspace{10mm}

\vspace{2mm}

\vspace{5mm}

\newpage

   \begin{center}
  {\huge  {\bf Appendix}} 
   \end{center}

\appendix

 \section{Rolling without twist:  curvature of an  Ehresmann connection}  \label{controlK}
\renewcommand{\theequation}{A.\arabic{equation}}

\setcounter{equation}{0}

  $\,\,\,\,$  Rolling without slip and twist are often considered in robot modeling   and control.  Optimization leads to interesting problems in subriemannian geometry \cite{Montgomerybook}. 
It has also been used  for doing data interpolation on manifolds and in  information geometry.

   Rubber rolling of two surfaces 
  has a   nice differential geometric interpretation:  \textit{at the   corresponding points the curves     have equal geodesic curvatures.}   
See  \cite{EhlersKoiller2007}  for an elementary proof.  
 The kinematics of two  rolling manifolds  without slip and twist was  developed  in high generality.
 
  For instance, in  Sharpe's book   (\cite{Sharpebook}, Appendix B)      two n-dimensional manifolds $\Sigma_1, \, \Sigma_2$ inside an   euclidean ambient are considered. 
  In view of Nash embedding theorem this extrinsic viewpoint suffices.  
An  intrinsic viewpoint  is  even more appealing. 

 Consider  two abstract manifolds of the same dimension\footnote{Different dimensions  can also be considered \cite{Mortada2015}.  One example is  the famous  rolling problem
of an infinitesimally  thin disk (1d) over the plane (2d), but now imposing also the no-twist condition.  Pilots  ``roll''  the airplane surface (2d) in space  (3d) in oder to control the adverse yaw. },  equipped with   Riemannian metrics $g_1$ and $g_2$. The configuration space consists of  points  of contact $x$ on the first, $\hat{x}$ on the second and a matrix  $A $ in $SO(n)$  identifying the tangent spaces  of  $T_x\Sigma_1$  with  $T_{\hat{x}} \Sigma_2$.

  The total  configuration space $C = C(\Sigma_1,\Sigma_2)$  is a  fiber bundle $$SO(n) \hookrightarrow C(\Sigma_1,\Sigma_2) \to \Sigma_1\times\Sigma_2 . $$ 
  
  A  local section can be constructed with two  orthonormal  frames $F_1, \, F_2$  for the corresponding metrics. The  identity matrix makes  $F_1 \mapsto F_2$,  so the set   $ \{ \, A: T_x \Sigma_1 \rightarrow T_{\hat{x}} \Sigma_1\, \}  \equiv SO(n)$. \\

   The no-slip condition is  just the statement that for corresponding curves $$\gamma_1:  \, x = x(t), \, \text{and}\,\, \gamma_2:  \, x = \hat{x}(t), \,\,\, \text{then} \,\,\, 
A(t)\dot{x}(t)=\dot{\hat{x}}(t) .
  $$
  The no-twist condition amounts  to the following requirement: 
  
   \textit{Any vector field that is parallel  along any given curve for the Levi-Civita connection of one  metric   corresponds, on the other curve,  to a vector field  that is parallel relative to the Levi-Civita connection of  the other metric.  } \\
  
These conditions together produce the admissible distribution of rank $n$ on $C$.  Choosing one of them, say $\Sigma_2$,   to ``roll'' on the other,  we  have an Ehresmann connection on $C \rightarrow \Sigma_2$.

Let   $n=2$.  Following   \cite{Agrachev1999},  or Bryant and Hsu (\cite{Bryant1993},  section 4.4), one takes local orthonormal frames  $(e_1(x),e_2(x))$  in $\Sigma_1$ and $(e_1(\hat{x}),e_2(\hat{x}))$  in $\Sigma_2$. 
Let $\phi $ the angle of rotation from the frame $ A e_1, A e_2$  to the frame $\hat{e}_1, \hat{e}_2$. 
 The admissible distribution is spanned by 
\begin{equation} \begin{split}
& X_1 = e_1+ \cos \phi \hat{e}_1 +\sin \phi \hat{e}_2 + (-\alpha_1 + \hat{\alpha}_1 \cos \phi + \hat{\alpha}_2 \sin \phi) \, \frac{\partial}{\partial \phi} \\
& X_2  = e_2 - \sin  \phi \hat{e}_1 +\cos \phi \hat{e}_2 + (-\alpha_2 -  \hat{\alpha}_1 \sin  \phi + \hat{\alpha}_2 \cos  \phi) \, \frac{\partial}{\partial \phi}
\end{split}
\end{equation}
where $\alpha_i, \hat{\alpha}_i$  are the  structure constants:
\begin{equation}  [e_1 ,  e_2] (x)  = \alpha_1(x) e_1  + \alpha_2(x) e_2\,\,  , \,\, [\tilde{e}_1 , \tilde{e}_2](\tilde{x}) = \tilde{\alpha}_1(x) e_1  + \tilde{\alpha}_2(x) e_2 .
\end{equation} 

One has
\begin{equation}
  [X_1 , X_2 ]  =  \alpha_1 X_1 + \alpha_2 X_2 + (\kappa(\tilde{x}) - \kappa(x)) \, \frac{\partial}{\partial \phi}
 \end{equation}
 so the  connection curvature  is the difference of the Gaussian curvatures, 
 \begin{equation}  \label{agrachevformula1}
  K = (\kappa(\tilde{x}) - \kappa(x)) \, \frac{\partial}{\partial \phi} \,\,\,\,\, \,  \text{(the sign depends on orientation choices).} 
 \end{equation}
 
Some more references are  \cite{Chitour2012},  
   \cite{An2014},  
\cite{Grong2012}, \cite{Grong2016}, and  specially the papers 
by Fatima Leite and her collaborators: \cite{Fatima2008},  \cite{FatimaAIP2009},  \cite{Fatima2011} \cite{Fatima2011a}, \cite{Fatima-intrinsic2012}, \cite{Fatima2015}.  There one can find       nice proofs  for the preservation of  geodesic curvature (in any dimension);  the paper 
\cite{Fatima2008}  discusses the so-called ``Riemann polynomials" and 
\cite{Fatima2015} , \cite{Jurdjevic2023}  studies  Riemannian symmetric spaces.
Rolling of manifolds of different dimensions have been considered since the `ancients'  \cite{appell1900integration} and was recently revisited \cite{Mortada2015}. \\

 \newpage 
 
  \section{$J \cdot K$  and $\Phi$-simple systems    (by  Luis Garc\'ia-Naranjo)} 
  \label{referee1}

\renewcommand{\theequation}{B.\arabic{equation}}
\setcounter{equation}{0}

The coeﬃcients $f^{ij}_k$ in equation (\ref{almostNH})   are very interesting quantities. They are the coordinate
representation of a skew-symmetric (1,2) tensor on  $Q$  that measures the interplay between
the constraint distribution and the kinetic energy metric. This tensor conveniently encodes
the geometry of nonholonomic Chaplygin systems in a way that is useful to analyze measure
preservation, Hamiltonization, etc. It appears in Koiller \cite{Koiller1992} and Cantrijn et a.l \cite{Cantrijn2002}  
and is denoted the gyroscopic tensor T in   \cite{Garcia-Naranjo2019, Garcia-Naranjo2020}. According to the conventions taken 
  in  \cite{Garcia-Naranjo2020}, one has
\begin{equation}
{\cal T}\left(\frac{\partial}{\partial r_i} , \frac{\partial}{\partial r_j} \right)  = - \sum_{k=1}^m \, f^{ij}_k \, \frac{\partial}{\partial r_k}
\end{equation}

In particular, considering that for the example treated in the paper one has,
\begin{equation} J \cdot K = - p_\psi \, n(\theta) \, d\theta \wedge d\psi, 
\end{equation}
we deduce that
\begin{equation}
f_\theta^{\theta \psi} = 0 \,,\,\,  f_\theta^{\theta \psi} = - n(\theta)
\end{equation}
and therefore, the gyroscopic tensor is determined by

\begin{equation}
{\cal T}\left(\frac{\partial}{\partial \theta} , \frac{\partial}{\partial \psi} \right)  = n(\theta) \, \frac{\partial}{\partial \psi}
\end{equation}
A fundamental property of this particular $ {\cal T } $ is that
\begin{equation}
{\cal T}(Y,Z) = Z[\Phi]Y-Y[\Phi]
\end{equation}
for any pair of vector fields $Y,Z$ on $S^2$ (where $Z[\Phi] = L_Z\Phi$ and similarly for $Y[\Phi]$), where
\begin{equation} \Phi  : S^2 \rightarrow \R, \,\,\, \Phi = - \int n(\theta) \,d\theta = - \log N(\theta)
\end{equation}
In the terminology introduced in \cite{Garcia-Naranjo2019, Garcia-Naranjo2020} this means that the system is ``$\Phi$-simple.
 
There are a number of consequences which follow from the $\Phi$-simple structure of the system:
Hamiltonization, measure preservation, and also, since $\psi$  is cyclic for the compressed Hamiltonian 
and 
$\partial \Phi/\partial \psi = 0 $  
 the existence and explicit form of the first integral $\ell$ (see [13, Theorem
3.1]).

 Furthermore, the recent paper \cite{Simoes2025}   
  shows that, up to a time reparametrization,
the trajectories of the unreduced dynamics coincide with the trajectories of an unconstrained
lagrangian flow on $Q= SO(3) \times \R^2 $ for which the rubber rolling distribution is invariant.

Essentially, $\Phi$-simple Chaplygin systems constitute the class of nonholonomic systems which
more closely resemble classical Hamiltonian systems on symplectic manifolds. Therefore, it is
 natural to expect that the second MW-type reduction that eliminates $\psi$  at constant values of
$p_\psi$ goes through.
To my knowledge, there is no general theorem that guarantees this on the
literature, but I do not think it is diﬃcult to  prove such a result along the lines of [13,
Theorem 3.1].

Do I understand correctly? If yes, the ``miracle"  referred to by ther author is that $\Phi$ has an explicit form
regardless of the body shape.

Futhermore, in the final commentrs the
author says 
\begin{quote}“We have the feeling that Chaplygin’s reducing multiplier method is related
to the $J \cdot K$ term and Hamiltonization, but we leave this just as loose question, up for grabs.” 
\end{quote}
This
question is partially settled: it is known that $\Phi$-simplicity is a suﬃcient condition for 
the application of Chaplygin’s reducing multiplier method  \cite{Garcia-Naranjo2019,Garcia-Naranjo2020,Dragovic2023}. There are
several multi-dimensional examples of Hamiltonization relying on this observation, like the Veselova system (see \cite{ Garcia-Naranjo2020}).\\

  \newpage
 
 {\footnotesize
 
 \bibliographystyle{unsrt}
\bibliography{rolling.bib}

}

\end{document}